\documentclass[pdflatex,sn-aps]{sn-jnl}

\usepackage{graphicx}%
\usepackage{multirow}%
\usepackage{amsmath,amssymb,amsfonts}%
\usepackage{amsthm}%
\usepackage{mathrsfs}%
\usepackage[title]{appendix}%
\usepackage{xcolor}%
\usepackage{textcomp}%
\usepackage{manyfoot}%
\usepackage{booktabs}%
\usepackage{algorithm}%
\usepackage{algorithmicx}%
\usepackage{algpseudocode}%
\usepackage{listings}%
\usepackage{mathtools}
\usepackage{lineno}%
\usepackage{import}
\usepackage{tikz}
\usetikzlibrary{math}
\usetikzlibrary{calc}
\usetikzlibrary{decorations.pathreplacing}

\usepackage[normalem]{ulem}
\usepackage[caption=false]{subfig}

\newcommand{\be}{\begin{equation}}
\newcommand{\ee}{\end{equation}}
\newcommand{\ba}{\begin{eqnarray}}
\newcommand{\ea}{\end{eqnarray}}

\newcommand{\Tr}{\mathrm{Tr}}
\newcommand{\jmax}{j_{\mathrm{max}}}

\begin{document}

\title[Entanglement in SU(2) Gauge Theory]{Entanglement Properties of SU(2) Gauge Theory}

\author[1]{\fnm{Lukas} \sur{Ebner}}\email{Lukas.Ebner@physik.uni-regensburg.de}
%\equalcont{These authors contributed equally to this work.}

\author[2]{\fnm{Berndt} \sur{M\"uller}}\email{bmueller@duke.edu}
%\equalcont{These authors contributed equally to this work.}

\author*[1]{\fnm{Andreas} \sur{Sch\"afer}}\email{Andreas.Schaefer@physik.uni-regensburg.de}
%\equalcont{These authors contributed equally to this work.}

\author[1]{\fnm{Leonhard} \sur{Schmotzer}}\email{Leonhard.Schmotzer@physik.uni-regensburg.de}
%\equalcont{These authors contributed equally to this work.}

\author[1]{\fnm{Clemens} \sur{Seidl}}\email{Clemens.Seidl@physik.uni-regensburg.de}
%\equalcont{These authors contributed equally to this work.}

\author[3]{\fnm{Xiaojun} \sur{Yao}}\email{xjyao@uw.edu}
%\equalcont{These authors contributed equally to this work.}

\affil*[1]{\orgdiv{Institut f\"ur Theoretische Physik}, \orgname{Universit\"at Regensburg}, \city{Regensburg}, \postcode{D-93040}, \country{Germany}}

\affil[2]{\orgdiv{Department of Physics}, \orgname{Duke University}, \city{Durham}, \postcode{27708}, \state{NC}, \country{USA}}

\affil[3]{\orgdiv{InQubator for Quantum Simulation, Department of Physics}, \orgname{University of Washington}, \city{Seattle}, \postcode{98105}, \state{WA}, \country{USA}}

\abstract{We review recent and present new results on thermalization of nonabelian gauge theory obtained by exact numerical simulation of the real-time dynamics of $(2+1)$-dimensional SU(2) lattice gauge theory. We discuss: (1) tests confirming the Eigenstate Thermalization Hypothesis; (2) the entanglement entropy of sublattices, including the Page curve, the transition from area to volume scaling with increasing energy of the eigenstate and its time evolution that shows thermalization of localized regions to be a two-step process; (3) the absence of quantum many-body scars when higher gauge field representations are taken into account; (4) the spectral form factor, which exhibits the expected slope-ramp-plateau structure for late times; (5) the entanglement Hamiltonian for SU(2), which has properties in accordance with the Bisognano-Wichmann theorem; and (6)  a measure for non-stabilizerness or ``magic'' that is found to reach its maximum during thermalization. We conclude that the thermalization of nonabelian gauge theories is a promising process to establish quantum advantage.}


\maketitle

\section{Introduction}\label{sec:Intro}

Euclidean lattice gauge theory (LGT) has made it possible to calculate many physical quantities in QCD {\it ab initio} with remarkable precision. Today, numerical simulations using powerful algorithms and high-performance computing platforms are providing theoretical control over increasingly complex properties of the QCD vacuum and its excitation spectrum, as well as the thermal properties of QCD matter. Unfortunately, these techniques do not apply to problems whose solution requires a formulation in Minkowski space, such as the dynamical response of systems governed by QCD to external perturbations. This is unsatisfactory, since systems of quarks and gluons out of equilibrium often thermalize on time and length scales that are much shorter than those in other systems, and thus it is interesting to understand their thermalization {\it ab initio}. For studies of thermalization mechanisms and time scales in a variety of quantum field theories, see \cite{Berges:2000ur,Micha:2004bv,Schlichting:2012es,Kurkela:2012hp,Ishii:2015gia, Guin:2025sts} and the review article \cite{Berges:2020fwq}.

The collision of two heavy ions in the ultra-high vacuum of a particle collider represents one of the most isolated quantum systems that can be investigated in the laboratory, and thus can serve as a prototype for the study of the apparent evolution toward (microcanonical) equilibrium in time-reversal invariant quantum systems. It is plausible that the fundamental mechanisms which are at work in high-energy heavy-ion collisions are closely related to those that have been understood to resolve the information problem of black hole physics \cite{Mathur:2009hf,Penington:2019kki,Almheiri:2019hni}. For QCD and nonabelian gauge theories, in general,  accurate numerical studies of real-time dynamics, thermalization and entanglement on classical and quantum computers are coming into reach~\cite{Pichler:2015yqa,Banuls:2019bmf,Dasgupta:2020itb,Davoudi:2022xmb,Bauer:2023qgm,Magnifico:2024eiy,Illa:2024kmf,Farrell:2024fit,Mueller:2024mmk,Calajo:2024qrc}.

It is believed that the equilibration of isolated quantum systems is closely linked to the dynamics of quantum entanglement, including the entanglement properties of its energy eigenstates. A powerful guiding principle in this context is the Eigenstate Thermalization Hypothesis (ETH) proposed by Joshua Deutsch and Mark Srednicki \cite{Deutsch:1991qu,Srednicki:1994mfb,DAlessio:2015qtq}. It posits that the matrix elements of generic local observables in the energy eigenbasis can be represented by a regular diagonal component with exponentially small fluctuations around the microcanonical expectation value, which are stochastic with a universal structure that resembles that of a Gaussian random matrix within sufficiently narrow energy bands. Although these fluctuations are exponentially suppressed, their large number allows them to dominate certain aspects of the dynamical behavior of the system. Thus, the structure of the matrix elements of energy eigenstates lies at the core of understanding thermalization in gauge theories. 

The ETH has been studied extensively in schematic models of strongly interacting quantum systems with a gravity dual, where the contributions from off-diagonal matrix elements have been shown to be dual to the contributions from Euclidean wormholes in the path integral \cite{Arkani-Hamed:2007cpn,Belin:2020hea}. As no exact gravity dual for QCD is known, this approach is not rigorously applicable to QCD. The evolution of a highly ordered (initial) state of an isolated interacting quantum system into a quasi-thermal state through its internal dynamics has been investigated for a wide range of model systems, from Ising-type spin chains, many-fermion systems, random matrix models, to low-dimensional gravity models   \cite{Winer:2023btb,Winer:2020mdc,Winer:2022hydrosff,Blake:2022uyo,Gogolin:2015gts,Nowak:2012gd,Hubeny:2013hz,Balasubramanian:2010ce,Xu:2022vko,Bardet:2021tdj,Dymarsky:2016ntg, Chen:2024lji,Jahn:2020ukq}. Various tools have been applied in these investigations including direct numerical integration of the Schr\"odinger equation on digital computers, multiscale entanglement renormalization on tensor networks (MERA), and holographic mapping to a gravity dual in Anti-de Sitter (AdS) space. 

The systematic study of this set of problems for nonabelian gauge theories and, specifically, QCD, has only just begun. The method of choice is the diagonalization of the Hamiltonian in a suitable truncated basis \cite{Byrnes:2005qx,Klco:2019evd}. This truncation is necessary because the Hilbert space is infinite. (In our simulation we typically construct up to $O(10^5)$ eigenstates of the Hamiltonian in the electric basis.) Naturally this choice is not unique. Other groups have used for SU(2) gauge theory a combination of Hilbert space truncation and regularization of the gauge group \cite{Haase:2020kaj}, q-deformed algebras
\cite{Zache:2023dko} or energy truncation in the Casimir operator basis \cite{Cataldi:2023xki,Zohar:2014qma}.
The simulation of the time evolution on a quantum computer has been performed in some simplified cases, but remains limited in precision due to constraints from the number and noisiness of available qubits. In this review, we focus on recent studies on digital computers which are limited to rather small systems because of the exponential growth of the Hilbert space with system size. In the future, such studies will serve as benchmarks for investigations on quantum computers, where the number of required error-corrected qubits only grows linearly with the system size.

As the complexity of quantum simulation is concerned, according to the Gottesman-Knill theorem \cite{Gottesman:1998hu}, stabilizer quantum circuits, which are circuits generated from only the controlled NOT, Hadamard, and phase gates, can be efficiently simulated in classical computers with polynomial resources, even though these circuits can generate high entanglement. Furthermore, the classical computing resource needed for a quantum circuit scales exponentially with the number of $T$ gates (also known as the $\pi/8$ gates)~\cite{Aaronson:2004xuh}. To describe the non-stabilizerness, a quantity called ``magic" $M_A\geq 0$  was invented, with a larger magic corresponding to a circuit with more $T$ gates. Therefore, only states with high and extensive magic 
demand quantum computing. Magic has been studied extensively, e.g., for low-energy nuclear systems~\cite{Robin:2024bdz,Brokemeier:2024lhq} and U(1) gauge theory~\cite{Falcao:2024msg}, and many relevant general results were obtained.

So far, numerical studies based on exact diagonalization of the lattice gauge theory (LGT) Hamiltonian have mainly focused on the gauge group SU(2). In spite of its inherent size limitations, enough progress has been made to answer some basic questions about entanglement and the validity of ETH for small lattices. The main thrusts in these studies concern the approach of a well-ordered initial state to thermal behavior, the approach of the entanglement entropy of a subsystem to the canonical value of the thermal entropy of the subsystem, and the entanglement structure of the final state. In addition to the review of already published results, we also present several new, previously unpublished results.

We now give an overview of the structure of this article. Existing numerical simulations studying the ETH related properties of (2+1)-dimensional SU(2) gauge fields are reviewed in Section~\ref{sec:ETH}. These also permit to explore the time scales on which random matrix properties are established. In Section~\ref{sec:SFF} we show that the spectral form factor for the SU(2) LGT exhibits the slope-ramp-plateau structure expected for a chaotic quantum system. The transition between the slope and ramp regions allows us to study the diffusive dynamics of the gauge theory before full thermalization. Section~\ref{sec:EE-Page} review results on the entanglement structure of subsystems, including the Page curve for the entanglement entropy as a function of subsystem size and the possible existence of quantum many-body scars (QMBS). In Section~\ref{sec:EE-time} we analyze the time evolution of Entanglement Entropy and its approach to the microcanonical limit. In Section~\ref{sec:EE-BW} we present an analysis of the entanglement Hamiltonian. Finally, in Section~\ref{sec:magic} we show first results for the anti-flatness of the entanglement spectrum and its time evolution. 
For the reader who is already familiar with the published literature, we point out that Fig.~\ref{fig:SFF_plots} and the associated Section \ref{sec:SFF}, Fig.~\ref{fig:evo_P9_J1_gsq1.2} and Section \ref{sec:EE-time}, Figs.~\ref{fig:BW-temp} and~\ref{fig:entanglement_spectra} in Section \ref{sec:EE-BW}, as well as Figs.~\ref{fig:thermal_magic} and \ref{fig:magic}, both in Section \ref{sec:magic}, present new results, while the remainder of the article is largely a review of previously published results.

\section{Methods}
\subsection{SU(2) Lattice Gauge Theory}
\label{sec:LGT}

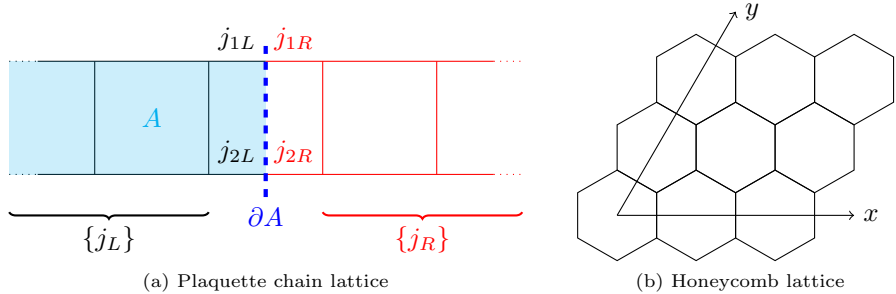
\begin{figure}[t]
\centering
\hfill
\subfloat[Plaquette chain lattice\label{fig:Plaquette chain}]{%
  \tikzmath{\l=1.5;\b=0.1;\t=0.3;}
    \begin{tikzpicture}
        
        \draw[] ($(0.5*\l, 0)$) -- ($(2.5*\l, 0)$);
        \draw[] ($(0.5*\l, \l)$) -- ($(2.5*\l, \l)$);
        \draw[dotted] ($(0.25*\l, 0)$) -- ($(0.5*\l, 0)$);
        \draw[dotted] ($(0.25*\l, \l)$) -- ($(0.5*\l, \l)$);
        
        % \draw[] ($(0, 0)$) -- ($(0, \l)$);
        \draw[] ($(\l, 0)$) -- ($(\l, \l)$);
        \draw[] ($(2*\l, 0)$) -- ($(2*\l, \l)$);
    
        \draw[red] ($(2.5*\l, 0)$) -- ($(4.5*\l, 0)$);
        \draw[red] ($(2.5*\l, \l)$) -- ($(4.5*\l, \l)$);
        \draw[red, dotted] ($(4.5*\l, 0)$) -- ($(4.75*\l, 0)$);
        \draw[red, dotted] ($(4.5*\l, \l)$) -- ($(4.75*\l, \l)$);
    
        \draw[red] ($(3*\l, 0)$) -- ($(3*\l, \l)$);
        \draw[red] ($(4*\l, 0)$) -- ($(4*\l, \l)$);
        % \draw[red] ($(5*\l, 0)$) -- ($(5*\l, \l)$);
    
        \path[fill=cyan,opacity=0.2] ($(0.25*\l, 0)$) -- ($(2.5*\l, 0)$) -- ($(2.5*\l, \l)$) -- ($(0.25*\l, \l)$) -- cycle;
        \node at ($(1.5*\l,0.5*\l)$) {$\color{cyan} A$};
    
        \draw[blue, ultra thick, dashed] ($(2.5*\l, -0.3)$) node[below]{$\partial A$} -- ($(2.5*\l, \l+0.3)$);
        
        \node at ($(2.25*\l,0)+(0,\t)$) {$j_{2L}$};
        \node at ($(2.25*\l,\l)+(0,\t)$) {$j_{1L}$};
        \node at ($(2.75*\l,0)+(0,\t)$) {$\color{red}j_{2R}$};
        \node at ($(2.75*\l,\l)+(0,\t)$) {$\color{red}j_{1R}$};
    
        \draw [
            thick,
            decoration={
                brace,
                mirror,
                raise=0.5cm
            },
            decorate
        ] ($(0.25*\l, 0)$) -- ($(2*\l, 0)$) 
        node [pos=0.5,anchor=north,yshift=-0.55cm] {$\{j_L\}$};
        \draw [red,
            thick,
            decoration={
                brace,
                mirror,
                raise=0.5cm
            },
            decorate
        ] ($(3*\l, 0)$) -- ($(4.75*\l, 0)$) 
        node [pos=0.5,anchor=north,yshift=-0.55cm] {$\color{red}\{j_R\}$};
        
    \end{tikzpicture}
}\hfill
\subfloat[Honeycomb lattice\label{fig:honeycomb}]{%
    \begin{tikzpicture}
        % draw hexagonal grid
        \def \s {0.6};  % length of hex side
        \def \cos {2*0.8660254037844387*\s};  % distance between hex centers
        
        \foreach \x in {0,...,2} {
            \coordinate (X) at ($(0,0)+(0:\x*\s*sqrt(3)$);
            \foreach \y in {0,...,2} {
                \coordinate (\x-\y) at ($(X)+(60:\y*\s*sqrt(3)$);
                \draw[] (\x-\y) +(-90:\s)
                \foreach \z [remember=\z as \lastz (initially 5)] in {0,...,5} {
                -- coordinate(\x-\y-\lastz-m) +(\z*60-30:\s) coordinate(\x-\y-\z)
                } -- cycle;
            } 
          }
          \draw[->] (0,0) -- (0:3*\cos)node[right]{$x$};
          \draw[->] (0,0) -- (60:3*\cos)node[right]{$y$};
          
    \end{tikzpicture}
    }
\hfill
\caption{
(a)
Plaquette chain segment in the vicinity of the cut separating the system into left (black) and right (red) parts, i.e., $A$ and $A^c$. The state on the left is labeled by the collection of $j$ values $\{j_L\}$ on the uncut links and the $j$ values $j_{1L}$, $j_{2L}$ for the two dangling links. Similarly for the state on the right. $j_{1L}=j_{1R}$ and $j_{2L}=j_{2R}$ since they form the same link before the cut.
(b)
Honeycomb lattice parallelogram with $N_x=N_y=3$ plaquettes.
}
\label{fig:plaquette_configurations}
\end{figure}

Motivated by the interest in real-time dynamics and the framework of eigenstate thermalization, we study pure SU(2) LGT in the Hamiltonian formulation.
For simplicity we consider a $(2+1)$-dimensional theory on square plaquette chains, as the one depicted in Fig.~\ref{fig:Plaquette chain}, and honeycomb lattices, as shown in Fig.~\ref{fig:honeycomb}. The advantage of these lattice configurations lies in the fact that all vertices have three links and can be expressed very efficiently by angular momentum coupling.\footnote{This was recently generalized to three dimensions using triamond lattices \cite{Kavaki:2024ijd}.} The Kogut-Susskind (KS) Hamiltonian for the spatially discretized gauge field on the plaquette chain takes the form \cite{Kogut:1974ag}
\begin{linenomath*}
\begin{align}
    \label{eq:H}
    H = \frac{g^2}{2}\sum_{\rm L} (E_i^a)^2 - \frac{2}{a^2g^2} \sum_{\rm P} {\rm Tr}\left[ \prod_{({\boldsymbol n},\hat{i})\in {\rm P}} 
    U({\boldsymbol n},\hat{i})\right] \,,
\end{align}
\end{linenomath*}
where $\rm L$ denotes links, $\rm P$ plaquettes, and the product in the second term is over the four oriented links of a plaquette. $U({\boldsymbol n},\hat{i})$ labels the Wilson line associated with the link at location $\boldsymbol n$ and direction $\hat{i}$.
$a$ in the denominator denotes the lattice spacing and $g$ is the gauge coupling.
The first term represents the electric energy of the gauge field; the second term encodes the magnetic energy.
For the honeycomb lattice the prefactors of the electric and magnetic terms in Eq.~\eqref{eq:H} change, cf. \cite{Muller:2023nnk} for more details.
Each link may be represented as a state $|j,m_L,m_R\rangle$ in the electric field basis \cite{Byrnes:2005qx, Zohar:2014qma, Liu:2021tef}, where $j=0,\frac{1}{2}, 1, ...$ denotes the angular momentum quantum number associated with the representation of SU(2) and $m_L$, $m_R$ are the left and right third component projections of $j$, respectively.
Physical states of the plaquette system are formed by the individual link states such that each vertex satisfies Gauss’s law, i.e., each vertex state is a color singlet. On lattices containing only sites with not more than three adjoining links, each physical state is represented unambiguously by the irreducible representations $j$ of the link variables. Since the plaquette operator itself is a singlet under the color gauge group, it can be shown that its action can be expressed solely in terms of the quantum numbers $j$ of the links that form the plaquette \cite{ARahman:2022tkr}.

The first term in Eq.~\eqref{eq:H} corresponds to a sum of local quadratic Casimir operators of SU(2). Hence, in the electric basis which labels the representation of each link, the electric energy $(E_i^a)^2$ is diagonal with eigenvalues $j(j+1)$.
The matrix elements of the second term may be expressed as a combination of Wigner-6$j$ symbols \cite{Klco:2019evd}. In order to obtain a finite-dimensional Hilbert space, the local Hilbert space of each link is truncated at $\jmax$. All matrix elements of the Hamiltonian for a given plaquette chain can then be calculated in the electric basis, and the Hamiltonian can be diagonalized numerically. In practice, this approach is limited by the exponential growth of the Hilbert space dimension with respect to the number of plaquettes and its polynomial growth with $\jmax$.\footnote{Since every link has at most $2\jmax+1$ possible values for $j$ and for the plaquette chain there are at most $N_L=3N_P+1$ unfixed links for both periodic and open boundary conditions with $N_P$ plaquettes, a rough upper bound for the dimension of the Hilbert space is given by $(2j_{\max} + 1)^{3N_P+1}$.} We have followed two different paths toward investigating the infinite volume limit for SU(2) properties: short plaquette chains with $\jmax>\frac{1}{2}$ as well as long chains and honeycomb lattices with $\jmax=\frac{1}{2}$. It is important to explore both approaches, as the $\jmax=\frac{1}{2}$ truncation may exhibit artifacts, such as quantum many-body scars, that are likely to disappear when higher electric field representations are included \cite{Ebner:2024mee}. Recent work claimed the existence of quantum many-body scars in $1+1$-dimensional U(1) and SU(2) and $(2+1)$-dimensional SU(2) lattice gauge theories with dynamical fermions, where gauge fluxes are similarly truncated and long-oscillating behavior is observed for certain initial conditions at certain couplings and masses~\cite{Halimeh:2022rwu,Calajo:2024bvs}, although it is not clear whether the bipartite entanglement entropies of most eigenstates at the studied couplings and masses form a narrow band, which is a necessary condition for the thermalization of most states and only weak ergodicity breaking. Unfortunately, our resources do not allow us to perform a controlled extrapolation to zero lattice spacing which is rather non-trivial, see e.g. \cite{Kluberg-Stern:1983lmr,Hasenfratz:1981tw}.

When we impose periodic boundary conditions on the plaquette chain, the KS Hamiltonian is translationally invariant and can be simultaneously diagonalized with the translation operator. Each eigenstate corresponds to a momentum sector, given by $p_k=2\pi k/N$, where $k\in[0,1, ..., N-1]$ and $N$ is the number of plaquettes. In the $k=0$ and $k=N/2$ sectors one can simultaneously diagonalize two parity operators corresponding to the two spatial dimensions of the system. These two momentum sectors of the Hamiltonian consist of four separate blocks corresponding to eigenvalues $\pm1$ of the two parity operators with no symmetries remaining.

If only states with $j=0,\frac{1}{2}$ are taken into account, the SU(2) gauge theory on the plaquette chain is equivalent to an Ising model with Hamiltonian \cite{Hayata:2021kcp,ARahman:2022tkr,Yao:2023pht}
\begin{linenomath*}
\begin{align}
H = \sum_{j=0}^{N-1} H_j \,, \quad H_j = \left( \frac{J}{2} (\sigma_j^z \sigma_{j+1}^z + \sigma_{j-1}^z \sigma_{j}^z) - 2J \sigma_j^z + h_x \nu_j \sigma_j^x \right) \, ,
\label{eq:H_Ising}
\end{align} 
\end{linenomath*}
where $J = -3g^2/16$, $h_x = (ag)^{-2}$, $\nu_j = (i/\sqrt{2})^{\sigma_{j-1}^z+\sigma_{j+1}^z}$. 
Periodic boundary conditions translate to $\sigma_{j+N}^z=\sigma_j^z$ in this formalism. 
The Ising mapping allows for numerical simplifications and efficient algorithms, making it possible to investigate a large number of plaquettes. The two paths described above enable calculations involving Hilbert spaces up to total dimension $10^6$.

Lastly, we briefly discuss the computation of entanglement entropy in the KS formulation of SU(2) gauge theory on the plaquette chain as presented in \cite{Ebner:2024mee}.
The entanglement entropy of a subsystem $A$ is defined as the von Neumann entropy $S_A=-{\rm Tr}(\rho_A\log\rho_A)$, where $\rho_A={\rm Tr}_{A^c} |\psi\rangle\langle\psi|$ denotes the reduced density matrix of the state on the segment $A$, and $A^c$ denotes the complement of $A$.
Each basis state on the plaquette chain can be written as a collection of $j$-quantum numbers of the individual links, i.e., $|\{j\}\rangle$.
Fig.~\ref{fig:Plaquette chain} depicts the vicinity of the boundary surface $\partial A$ of the subsystem $A$ which consists of all links to the left of $\partial A$. The system is thus divided by ``cutting'' through two horizontal links, which we now refer to as dangling links. The quantum state can be decomposed in terms of the left and right parts of the system, i.e., subsystem $A$ and its complement $A^c$,
\begin{linenomath*}
\begin{align}
    |\psi\rangle=c_{\{j_{L}\}j_{1L}j_{2L}\{j_{R}\}j_{1R}j_{2R}}\delta_{j_{1L}j_{1R}}\delta_{j_{2L}j_{2R}}|\psi_{\{j_{L}\}j_{1L}j_{2L}}\rangle\otimes|\psi_{\{j_{R}\}j_{1R}j_{2R}}\rangle\,,
\end{align}
\end{linenomath*}
where $c$ represents the combinatorial coefficients and we assume Einstein's summation convention. The Kronecker symbols ensure that the representation of the left and right dangling link variables  coincide.
Tracing out $\{j_{R}\}$, $j_{1R}$, $j_{2R}$, i.e., the physical degrees of freedom of the right segment, yields a reduced density matrix
that is a direct sum of density matrices associated with specific dangling link configurations. Since the separation of the subsystem along the horizontal dangling links does not involve vertices, Gauss's law is still valid at each vertex ensuring gauge invariance of the reduced density matrix.
It has been shown that ${\rm Tr}(\rho_A^n)={\rm Tr}(\rho_{A^c}^n)$ for any integer $n\geq2$ \cite{Ebner:2024mee}, implying equality of the eigenvalues of $\rho_A$ and $\rho_{A^c}$ and thus $S_A=S_{A^c}$.

\section{Results}
%\subsection{Thermalization}
%\label{sec:Thermal}

\subsection{Eigenstate Thermalization Hypothesis}
\label{sec:ETH}

One promising idea of how thermodynamics can be expressed in the framework of quantum mechanics is the ETH  \cite{Deutsch:1991qu,Srednicki:1994mfb}. It states that the structure of a local observable expressed in the energy eigenbasis takes the following form:
\begin{linenomath*}
\begin{equation}
\langle E_\alpha |{\cal A}| E_\beta \rangle = \langle {\cal A} \rangle_{\rm mc} (E)\delta_{\alpha\beta} + e^{-S(E)/2} f_{\cal A}(E,\omega) R_{\alpha\beta} \,,\!
\label{eq:ETH}
\end{equation}
\end{linenomath*}
where $E=(E_{\alpha}+E_{\beta})/2$ and $\omega=E_{\alpha}-E_{\beta}$. The first part of the formula only contributes to the diagonal part of the matrix, where $\langle \mathcal{A} \rangle_{\rm mc}(E)$ is the microcanonical ensemble value of the observable $\mathcal{A}$ at the energy $E$. The second part of the formula contributes to each element of the matrix and has an exponential suppression factor with respect to the microcanonical entropy $S(E)$ that is proportional to the size of the system. The $R_{\alpha \beta}$ are pseudo-random numbers with zero mean and unit variance for $\alpha \neq \beta$ and variance 2 for $\alpha=\beta$. If a sufficiently small energy window is considered, these $R_{\alpha \beta}$ should behave as drawn from a Gaussian Orthogonal Ensemble (GOE) and thus represent a universal component which, according to the hypothesis, should hold for every thermal system. $f_{\cal A}(E,\omega)$ is the spectral response function and encodes information about the dynamics of the system when it is interrogated by the observable $\cal A$. $f_{\cal A}(E,\omega)$ is related to the nonequal-time correlation function via a Fourier transform~\cite{DAlessio:2015qtq}.

ETH is expected to hold for highly excited states (i.e., not for the ground or low-lying states) of a Hamiltonian that is quantum chaotic. According to the Bohigas-Giannoni-Schmit conjecture, a time-reversal invariant Hamiltonian exhibits quantum chaos if the energy level statistics follows that of GOE \cite{Bohigas:1984characterization}. If all highly excited states of a quantum chaotic Hamiltonian fulfill ETH, the system is said to obey strong ETH. This can be weakly violated by quantum many-body scars (QMBS), e.g., excited nonthermal states. The latter possibility will be addressed in Section \ref{sec:EE-Page}.

To obtain results for the full (2+1)-dimensional KS Hamiltonian the limit of two parameters, the cutoff $\jmax$ and the number of plaquettes for each dimension $N_x,N_y$, must be taken to infinity. Although this is out of reach with exact diagonalization, we hope that even small systems can already give an understanding of the qualitative behavior of the gauge theory with respect to ETH.\footnote{
Although the classical gauge field action is scale invariant, quantized SU($N$) gauge fields develop a trace anomaly, acquire a nonperturbative mass gap, and their vacuum state exhibits a short correlation length related to the confinement scale. At temperatures $T$ above the confinement scale, all degrees of freedom are characterized by finite correlation lengths scaling with $1/T$ (in the perturbative domain, these can be of order $(gT)^{-1}$ or $(g^2T)^{-1}$). We expect this behavior to persist also for highly excited energy eigenstates. This expectation is confirmed by our numerical results, For example, for a chain of 17 plaquettes and $\jmax=\frac{1}{2}$, the correlation length is $l_{corr} \lesssim 0.8$, which means that regions of the lattice separated by at least two lattice constants do not strongly affect one another. Thus small systems should already exhibit behavior similar to that of large lattices.}

Here, we will consider two systems that maximize either the cutoff or size parameter, constrained by our resources, namely a honeycomb lattice of 20 plaquettes ($N_x=5, N_y=4)$ with minimal nontrivial cutoff $\jmax=\frac{1}{2}$ and a chain with three plaquettes with cutoff $\jmax=4$, where we choose an energy window in which the energy eigenvalues are converged. For a detailed analysis of another system, a long chain of 19 plaquettes and $\jmax=\frac{1}{2}$, see~\cite{Ebner_2024}.

One question that has not yet been answered precisely is for which operators the ETH is applicable. It makes sense to restrict this question to physically relevant operators that involve a limited number of degrees of freedom. Specifically for a gauge theory, operators should be further restricted to be gauge invariant, multiplicatively renormalizable, and have local support. So far, numerical investigations have focused on operators involving the total electric energy or Wilson loops around one plaquette or two adjacent plaquettes.

\begin{figure}[t!]
\subfloat[\label{fig:r_P3_jmax4_g2_0.6_converged_window}]{%
  \includegraphics[width=0.5\linewidth]{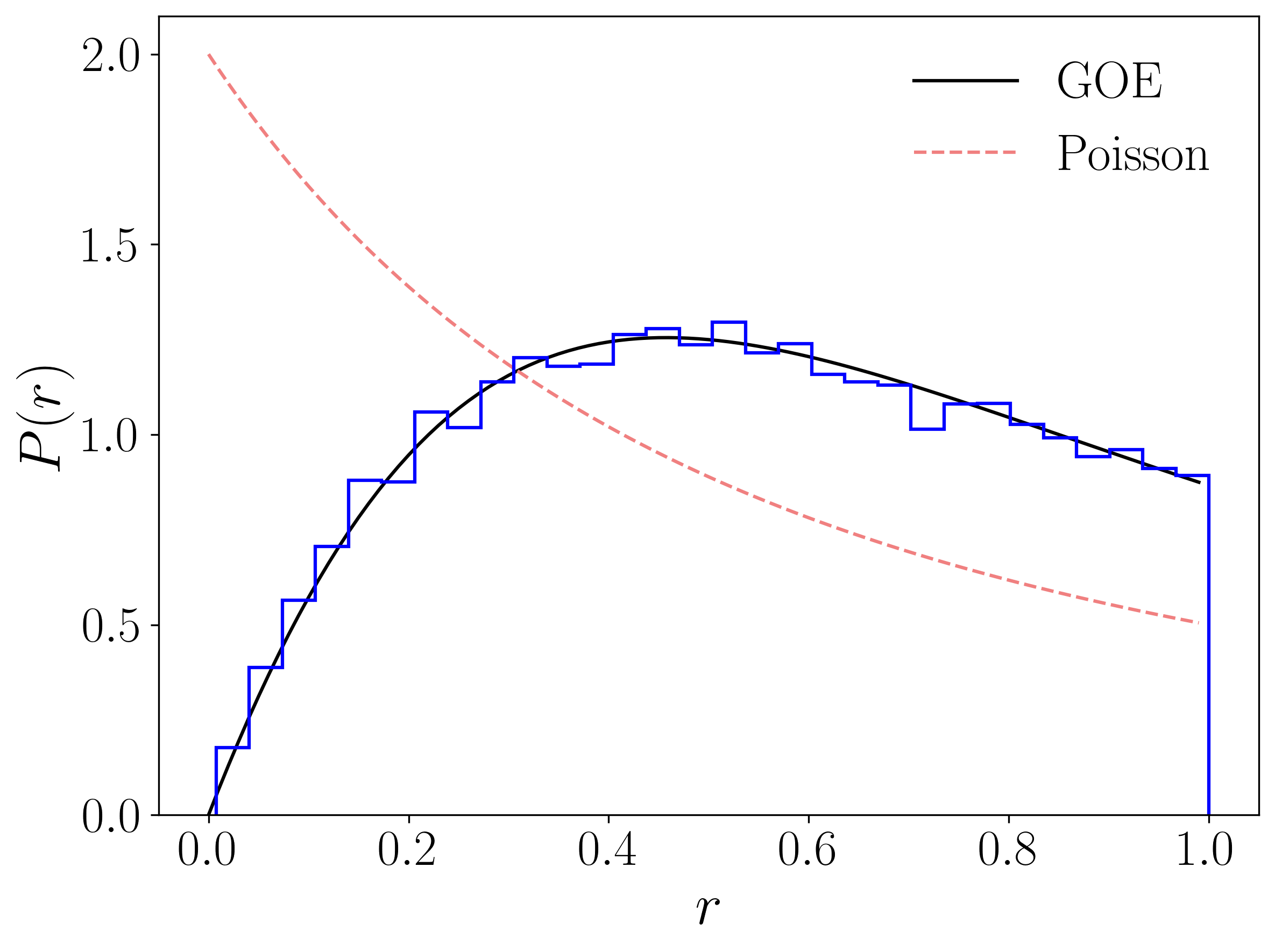}%
}\hfill
\subfloat[\label{fig:diag_periodic}]{%
  \includegraphics[width=0.5\linewidth]{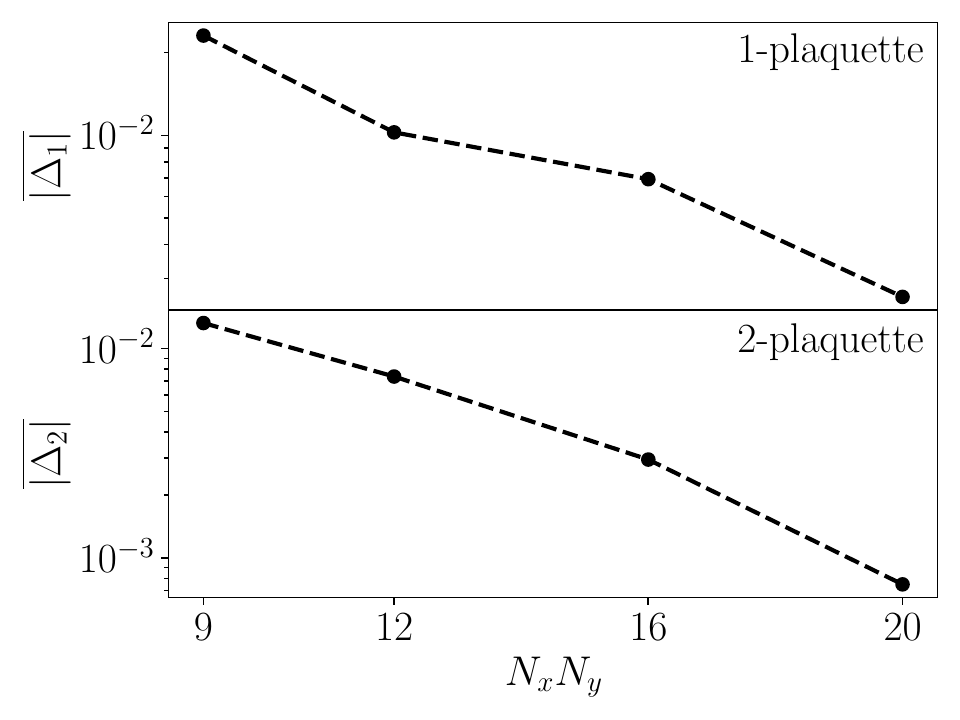}%
}
\caption{(a) Histogram of the restricted gap ratio (blue) compared to the GOE (black) and the Poisson (dashed red) prediction for the converged region of the spectrum of the plaquette chain with $N=3$, $ag^2=0.6$ and $\jmax=4$~\cite{Ebner_2024}. (b) Mean fluctuations $|\Delta_i|$ of the $i$-plaquette Wilson loop operator around the microcanonical value for different system sizes $N=9,12,16,20$ of the honeycomb lattice with $\jmax=\frac{1}{2}$, $ag^2=1$ and all momentum sectors up to $k_x=\lfloor N_x/2 \rfloor$, $k_y=\lfloor N_y/2 \rfloor$~\cite{Ebner_2024}.}
\label{fig:gap_diagonal}
\end{figure}

In order to compare our results to Random Matrix Theory (RMT) predictions for periodic boundary conditions, we must select one symmetry sector of the block diagonal Hamiltonian. We will perform the analysis of the energy level statistics for the plaquette chain in the sector of reflection symmetric states with momentum $k=0$, and for the honeycomb lattice in the sector with momentum $k_x=k_y=1$. A useful measure of the distribution of energy levels is the restricted gap ratio $r_\alpha$, which is defined as~\cite{Oganesyan_2007}
\begin{equation}
    0 < r_\alpha = \frac{{\rm min}[\delta_\alpha,\delta_{\alpha-1}]}{{\rm max}[\delta_\alpha,\delta_{\alpha-1}]} \leq 1 \, ,
\label{eq:rgap}
\end{equation}
where $\delta_\alpha = E_{\alpha+1} - E_\alpha$ is the difference between two adjacent energy eigenvalues.
This is a convenient choice, because $r_\alpha$ is independent of the local density of states. We compare the distribution of $r_\alpha$ of our system with the GOE prediction for a $3\times3$ matrix
%\begin{equation}
%    P_{\rm GOE}(r) = \frac{27}{4} \frac{r+r^2}{(1+r+r^2)^{5/2}}
%\label{eq:rgap_GOE}
%\end{equation}
and its mean value $\langle r \rangle_{\rm GOE}\approx0.5307$, obtained numerically for large matrices. The prediction for uncorrelated eigenstates is the Poisson distribution with mean value $\langle r \rangle_{\rm PO}\approx0.38629$~\cite{Atas_2013}.

The Hamiltonian (\ref{eq:H}) has two integrable limits: $ag^2 \rightarrow \infty$ and $ag^2 \rightarrow 0$. At intermediate coupling ($ 0.01 \lesssim ag^2 \lesssim 1.2 $), the eigenvalue distribution of our small lattice system follows accurately the GOE prediction and the system exhibits thus by definition quantum chaos. Outside this range the level spacing statistics starts to deviate from GOE and approaches a Poisson-like distribution. This happens because either the electric or the magnetic term begins to dominate, and both on their own are integrable. A good parameter choice for the plaquette chain is $ag^2=0.6$ with $\langle r \rangle=0.5329$ as shown in Fig.~\ref{fig:r_P3_jmax4_g2_0.6_converged_window} and for the honeycomb lattice $ag^2=1$ with $\langle r \rangle=0.5278$~\cite{Ebner_2024}. 

\begin{figure}[t!]
\subfloat[\label{figs:f_function_log_log_P3_jmax4_g2_0.6_converged_window}]{%
  \includegraphics[width=0.5\linewidth]{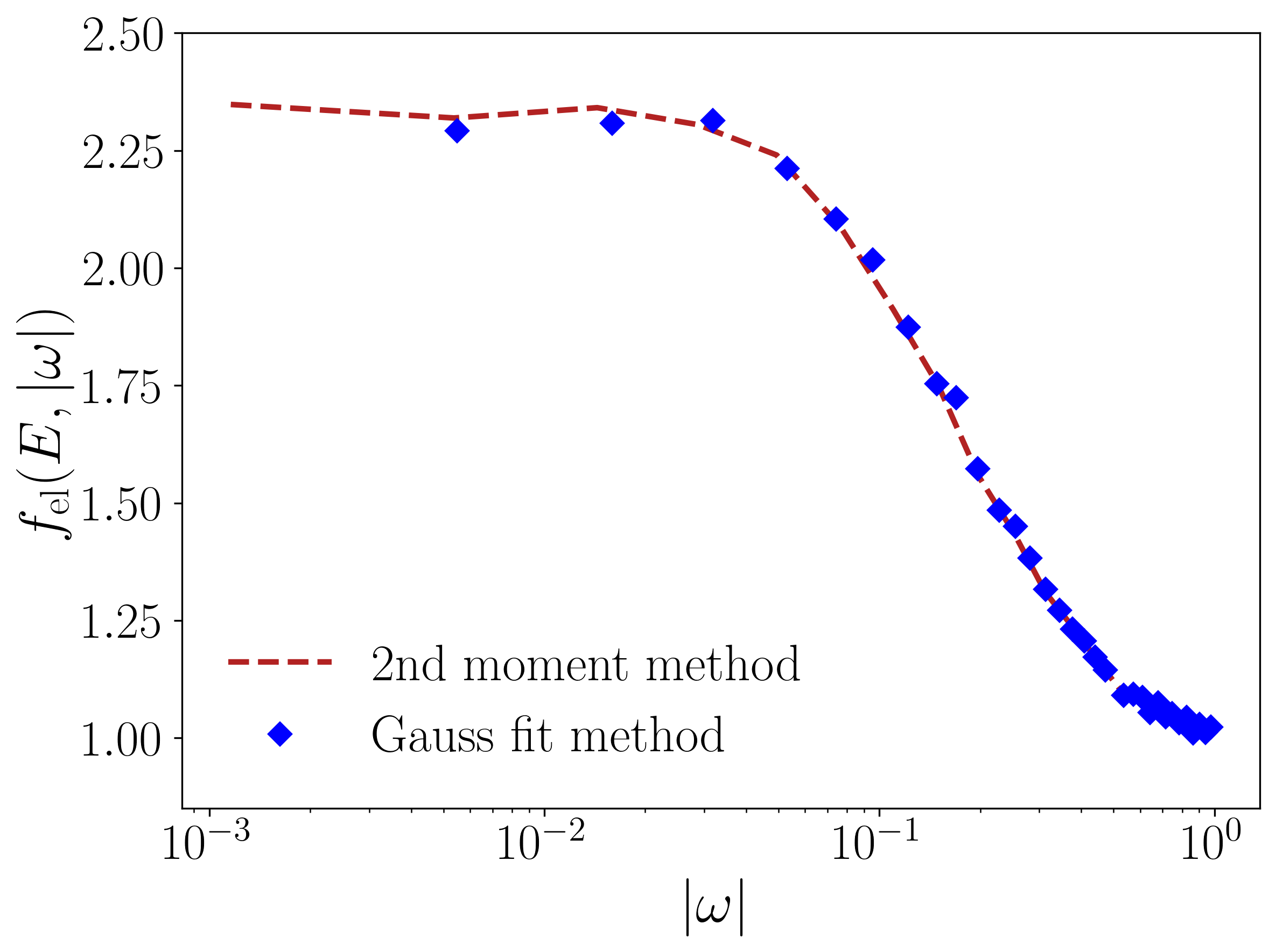}%
}\hfill
\subfloat[\label{figs:Lambda_T_Hel_P3_jamx4_g2_0.6_Delta0.2}]{%
  \includegraphics[width=0.5\linewidth]{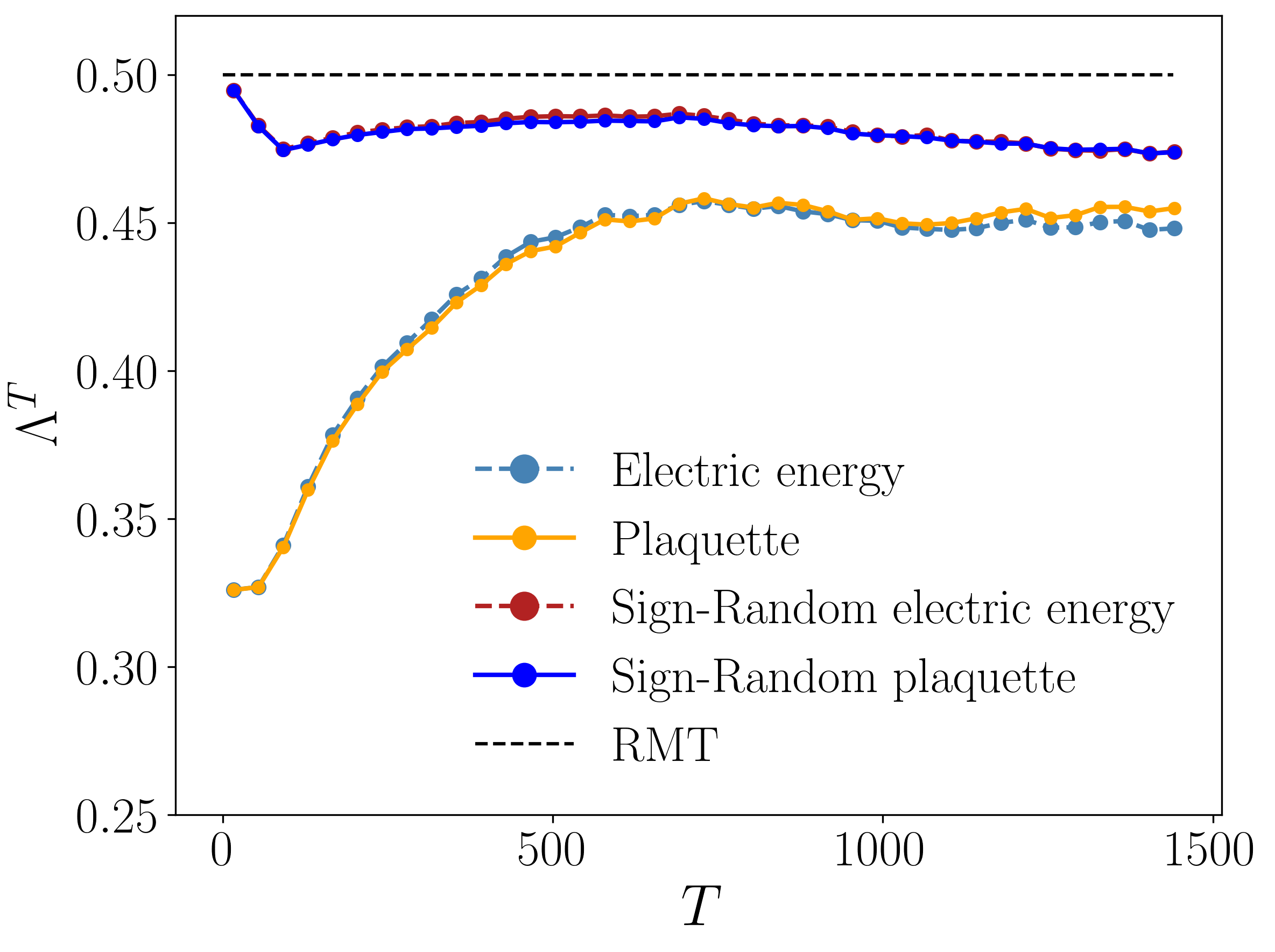}%
  }
\caption{(a) Double logarithmic plot of the small $\omega$ region of the spectral function $f_{\mathrm{el}}(E,\omega)$ calculated via the second moments (red) of the matrix elements and the Gaussian fit method (blue) for a system with $N=3$ plaquettes, $\jmax=4$ and $ag^2=0.6$. (b) $\Lambda^T$ measure as a function of $T$ for the total electric energy operator, the plaquette operator and their sign randomized versions, averaged over 100 samples, in the energy window around $E=23.5$ with width $\Delta E=0.1$ for a system with $N=3$ plaquettes, $\jmax=4$ and $ag^2=0.6$.}
\label{fig:f_lambda}
\end{figure}

To test the ETH, we first look at the diagonal part of the observable expressed in the energy eigenbasis. We verify that the expectation values $\langle E_\alpha | \mathcal{O} | E_\alpha \rangle $ depend smoothly on the energy $E_\alpha$. Therefore, we can define the microcanonical value for a certain energy $E_\alpha$ as $\langle \mathcal{O} \rangle _{\rm mc}(E_{\alpha})=\frac{1}{21}\sum_{\beta=\alpha-10}^{\alpha+10} \langle \beta | \mathcal{O} | \beta \rangle$. With this we can calculate the deviation of a diagonal matrix element from the microcanonical value $\Delta_i(\alpha)=\langle \alpha |O_i|\alpha \rangle-\langle \mathcal{O} \rangle _{\rm mc}(E_\alpha)$, take the mean over such fluctuations at different energies (except the edges of the spectrum) and compare results from different system sizes. According to the ETH ansatz, this quantity should be proportional to the factor $e^{-S(E)/2}$, which vanishes exponentially in the thermodynamic limit. In Fig.~\ref{fig:diag_periodic} this analysis is carried out for the one- and two-plaquette Wilson loop operators on the honeycomb lattice~\cite{Ebner_2024}. It can be clearly seen that the fluctuations in the thermodynamic limit approach zero exponentially, as expected. 

Next we turn to the off-diagonal matrix elements of the total electric energy operator.
With these we calculate the spectral function in the energy window of width $\Delta E=0.05$ around $E=18$ on the three-plaquette chain, where all energy eigenvalues are converged with respect to $\jmax$. At first we focus on the small $\omega$ region. There are two ways of doing it. One is to calculate the square root of the second moments of the matrix elements in small $\omega$ bins. Another method using the ETH ansatz gives $f_{\rm el}(E,\omega)=\sqrt{\rho(E)} \sigma$, where $\sigma$ is the Gaussian width of the distribution of the matrix elements within a small $\omega$ bin and $\rho(E)$ is the state density. Fig.~\ref{figs:f_function_log_log_P3_jmax4_g2_0.6_converged_window} shows that both approaches give the same answer for the plaquette chain and thus confirm ETH. The same holds true for the honeycomb lattice~\cite{Ebner_2024}. Furthermore, this small $\omega$ region is characterized by a diffusive transport peak with a plateau. For $\omega \gtrsim 4$ the spectral function falls off exponentially. 

In order to test how well the statistical distribution of the matrix elements show RMT behavior in the small $\omega$ limit, we look at another GOE measure, $\Lambda^T$, defined as the ratio of the squared second moment to the fourth moment of the traceless part of the matrix~\cite{Wang:2021mtp} obtained by retaining only matrix elements $\mathcal{O}_{\alpha\beta}$ between eigenstates satisfying $|E_\alpha - E_\beta| \leq 2\pi/T$. $T$ therefore corresponds to a time scale and controls which matrix elements contribute, i.e. increasing $T$ results in a decreasing bandwidth of non-zero matrix elements. The GOE prediction is $\Lambda^T = 0.5$. Figure~\ref{figs:Lambda_T_Hel_P3_jamx4_g2_0.6_Delta0.2} shows the $\Lambda^T$ measure for the total electric energy operator and the plaquette operator in the converged region for an energy window of size $\Delta E=0.2$ around $E=23.5$. The measure is seen to increase with $T$ reaching a maximum of $0.47$ for the total electric energy. The fact that $\Lambda^T$ gets closer to the GOE limit, but does not fully reach the value $0.5$ indicates that the off-diagonal matrix elements of the operator still retain some correlations even within a very narrow energy window. We conjecture that this failure to completely reach the RMT limit is due to the small lattice size. The results for these two operators are almost identical. This is not surprising, as the off-diagonal elements of the electric energy and the magnetic energy are the same up to a minus sign. The prefactor that distinguishes the magnetic energy and the plaquette term makes no difference here, since we are looking at a ratio. This equality is confirmed by our calculations.
The only distinction is the diagonal term, from which we have subtracted the average diagonal value. This could be the reason why there is a small difference at late times.

In order to study the nature of the correlations in more detail, we sign-randomize both operators, i.e., we multiply the off-diagonal matrix elements randomly with $+1$ or $-1$ while keeping the matrix symmetric, and evaluate the $\Lambda^T$ measure again. Finally, we average over 1000 randomized matrices. We observe that the sign-randomized versions are much closer to the Gaussian limit $\Lambda^T=0.5$. This leads us to conclude that the remaining correlations reside in the signs of the matrix elements in the energy basis rather than in their absolute values.

\subsection{Spectral Form Factor}
\label{sec:SFF}

The Spectral Form Factor (SFF) is another useful and widely used tool to investigate the dynamical properties of quantum systems. It has been successfully applied in various contexts, from chaotic and disordered quantum systems to many-body quantum systems, statistical mechanics, and quantum gravity, as it provides insight into universal properties of quantum systems that exhibit ergodic dynamics. In particular, the SFF is a well-established measure for identifying signatures of RMT behavior \cite{Mehta:2004random, Bohigas:1984characterization,Guhr:1998random,Saad:2019sykgravity}. Furthermore, it has been shown by Winer and Swingle \cite{Winer:2020mdc} that the period immediately preceding the dominance of RMT dynamics is characterized by diffusive hydrodynamics, which encapsulates the slowest relaxation to thermal equilibrium. The timescale for the onset of hydrodynamics, sometimes called ``hydrodynamization'', is expected to be similar for all observables that do not involve other locally conserved quantities. Thus the SFF provides a valuable tool for our exploration of SU(2) LGT, which complements the array of methods and GOE measures discussed above.

The SFF is defined as the ensemble-averaged squared magnitude of the Fourier transform of the filtered two-point correlation function of the density of states $\rho(E) = \sum_{i} \delta(E - E_i)$ \cite{Winer:2022hydrosff}:
\begin{linenomath*}
\begin{equation}
\text{SFF}(T, f) = \overline{\sum_{i,j} f(E_i) f(E_j) e^{i(E_i - E_j)T}} =  \overline{\left|\int_{-\infty}^{\infty} dE f(E) \rho(E) e^{-iET} \right|^2} \, , \label{eq:SFF}
\end{equation}
\end{linenomath*}
where $E_n$ are the eigenvalues of the system, $T$ is the rescaled time,  $f(E)$ is a (Gaussian) filter function applied to select different regions of the spectrum, and the overline indicates the average over the random matrix ensemble.

\begin{figure}[htb]
\subfloat[\label{fig:SFF_ramp}]{%
  \includegraphics[width=0.49\linewidth]{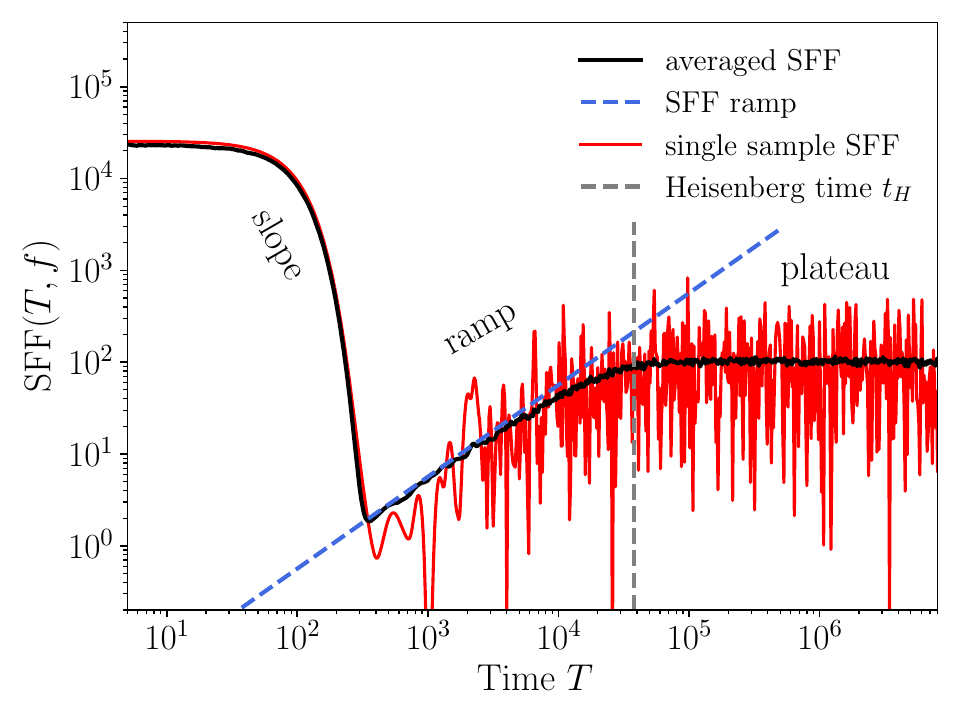}%
}\hfill
\subfloat[\label{fig:SFF_hydro}]{%
  \includegraphics[width=0.49\linewidth]{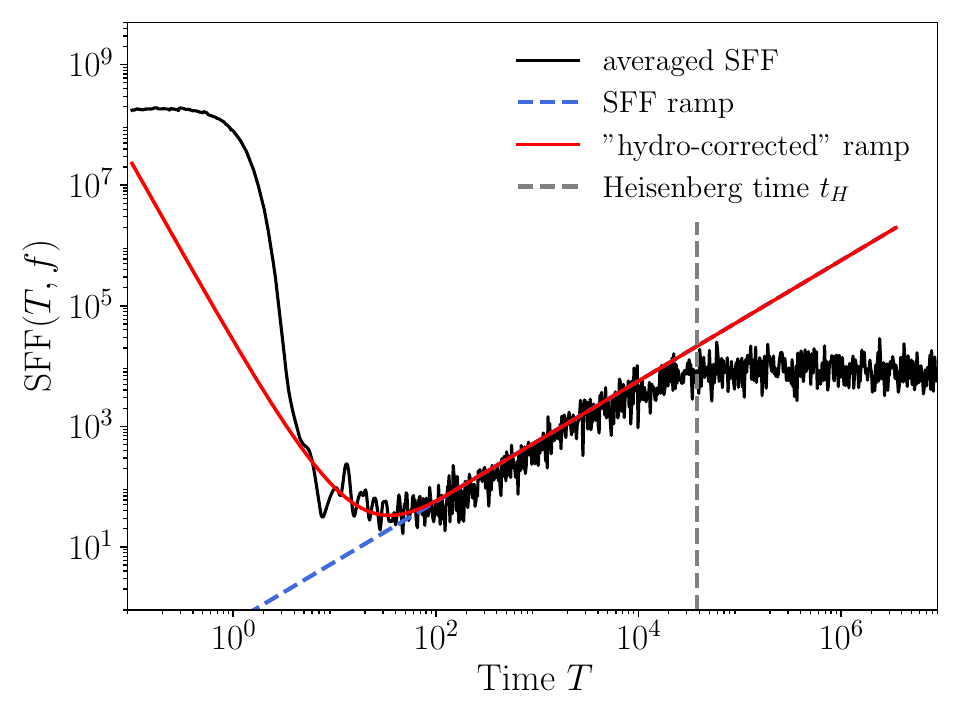}%
}
    \caption{(a) Averaged SFF (black) over 500 Gaussian windows with each $\sigma=0.01$ and the SFF of single sample (red) at $\mu=14$ with the same $\sigma$. The linear ramp (blue) is given in Eq. (\ref{eq:SFF_ramp}). (b) Averaged SFF (black) over 100 Gaussian windows with each $\sigma=1$ and fitted to it the "hydro-corrected" ramp (red) of Eq. (\ref{eq:SFF_hydro}) and linear ramp (blue). The systems in (a) and (b) are five-plaquette chains with $\jmax=\frac{3}{2}$, $ag^2=1.05$ and open boundary conditions $j=1,1,1,0$. The sampling of the Gaussian means is performed between $E=[10,22]$ (roughly the FWHM), which for this region of the spectrum results in a Heisenberg time $t_H/a = 3.818 \times 10^4$ (gray).}
    \label{fig:SFF_plots}
\end{figure}

The structure of the SFF for a random matrix ensemble can be divided into three characteristic regimes: the \emph{slope}, \emph{ramp}, and \emph{plateau} regions. The slope occurs at early times when the SFF is primarily influenced by the disconnected part of the two-point correlation function of the level density $\rho$. After the system has reached its Thouless time $t_\mathrm{Th}$ and fully explored its phase space, it shows ergodic behavior. This results in a linear ramp that is a universal characteristic for all chaotic systems described by RMT and given by
\begin{linenomath*}
\begin{equation}
    \text{SFF}_{\text{ramp}}(T,f) = \frac{T}{\pi \mathfrak{b}} \int dE\,  f^2(E)\,,
    \label{eq:SFF_ramp}
\end{equation}
\end{linenomath*}
where $\mathfrak{b}$ is the Dyson index of the Gaussian ensemble. The ramp ends when the system has reached its Heisenberg time $t_{H}=2\pi/\overline{\Delta E}$, where $\overline{\Delta E}$ is the mean level spacing. Beyond this point, the off-diagonal terms in Eq.~\eqref{eq:SFF} average to zero, and the SFF flattens into a plateau whose height is determined by the total number of eigenstates.

While in ETH the RMT behavior of generic operators is expected in the energy eigenstate basis within sufficiently small energy windows, the SFF probes for the randomness properties of the energy spectrum itself. These two properties are thought to be closely related and, as we have already demonstrated ETH behavior in the SU(2) LGT, we expect that its SFF also exhibits RMT behavior at late times. Since we only have one Hamiltonian for given system parameters, we replace the ensemble average with an average over many narrow energy windows roughly inside the Full Width at Half Maximum (FWHM) region of the spectrum.\footnote{This choice avoids artifacts from the tails of the spectrum.} This is realized by choosing a different center $\mu$ for each Gaussian filter function $f(E)=\exp{[-(E-\mu)^2/2\sigma^2]}$, which smooths out the fluctuations in the late-time SFF but does not affect the ramp because Eq. (\ref{eq:SFF_ramp}) is independent of $\mu$. In Fig.~\ref{fig:SFF_ramp} one can clearly see the expected RMT behavior at late times in the SFF of the SU(2) LGT.

If one chooses slightly broader energy windows, the averaging retains aspects of the structure of the physical Hamiltonian and captures the contributions from slow hydrodynamic modes. Such modes describe the transport of locally conserved quantities, such as energy and momentum or conserved particle number, which can only equilibrate by physical transport across distances of the order of the system size.\footnote{Although gauge symmetries also encode local conservation laws, they generally do not result in hydrodynamic transport behavior, as their associated collective transport manifests itself in gapped and strongly damped plasmon modes.} These modes lead to the appearance of a transition regime between the slope and the ramp regions of the SFF, which is absent for a pure random matrix ensemble. This hydrodynamic transition regime can be approximated by a ``hydro-corrected'' SFF ramp given by
\begin{linenomath*}
\begin{equation}
    \text{SFF}_{\text{hydro-ramp}}(T,f) =  \frac{T}{\pi \mathfrak{b}} \left[\prod_{\lambda \in \mathrm{spec}(-\Delta)} \frac{1}{1 - e^{- D \lambda T}} \right] \int dE\, f^2(E)\,, 
    \label{eq:SFF_hydro}
\end{equation}
\end{linenomath*}
where $\lambda \geq 0$ are the eigenvalues of the negative Laplacian $-\Delta$ and $D$ is the diffusion constant of the system (see Eqs.~(15) and (30) in \cite{Winer:2022hydrosff}). For a one-dimensional lattice of length $L$ with lattice spacing $a$, the eigenvalues are given by $\lambda_n = \left[2/a \sin\left(k_n a/2\right) \right]^2 $, where $k_n = 2 \pi n/L$ is the quantized momentum and $n\in \mathbb{Z}$ is constrained by $-L/2a < n \leq L/2a$ \cite{Kittel:2004intro}. We exclude the $k=0$ mode as it represents a spatially uniform contribution that fills the entire volume and, by nature, does not participate in the diffusive dynamics of the system.

As shown in Fig.~\ref{fig:SFF_hydro}, we are able to fit the ramp and its hydrodynamic precursor region in the five-plaquette chain system using Eq.~\eqref{eq:SFF_hydro} with $D/a= 0.0316 \pm 0.0006$. The closely related Thouless time $t_{\text{Th}}$ is dominated by the slowest hydrodynamic mode, identified by the smallest non-zero eigenvalue $\lambda_1$, since this mode sets the longest relaxation time for diffusive processes. Higher modes decay more rapidly, leaving the slowest mode as the primary contributor to the transition time toward the pure RMT behavior (see Eq.~\eqref{eq:SFF_hydro}). Thus, we find the Thouless time as $t_{\text{Th}}/a \sim (D\lambda_1)^{-1} = 22.9\pm0.4$.  This value matches closely with the onset of the ramp. When applied to larger lattices, this type of analysis will make it possible to compute real-time quantities of physical relevance, such as the diffusion constant and possibly other transport coefficients in the hydrodynamic regime.

%\section{Entanglement Properties}
%\label{sec:EE}

\subsection{Entanglement Entropy and Page Curve}
\label{sec:EE-Page}

We next study entanglement entropy as a function of the subsystem size for the ground state as well as for excited states on the plaquette chain configuration. Such investigations are very common whenever the apparent  thermalization of quantum states is discussed, and the resulting curves are generically called ``Page curves". Page originally introduced this notion in the context of random quantum states \cite{Page:1993df} and then applied it specifically to resolve the information problem of black holes \cite{Page:1993df}. The idea of the Page curve is that, measured by its entanglement entropy, any subsystem of a large quantum system that is in a randomly chosen pure quantum state appears thermal if the subsystem is smaller than half the full system. When the subsystem becomes larger than half the complete system, the entanglement entropy decreases until it reaches zero when the full system is considered. The presence of this behavior has been demonstrated for many chaotic quantum systems and confirmed experimentally in various experiments, see, e.g., Fig. 4 in \cite{Kaufman:2016mif}.

First results for the Page curve of periodic plaquette chains with electric field truncation $\jmax=\frac{1}{2}$ and aperiodic chains with $\jmax=1$ were presented in \cite{Ebner:2024mee}. In the following, we extend this work by investigating a periodic nine-plaquette chain with $\jmax=1$ in the zero-momentum and reflection symmetric sector.

The calculation of entanglement entropy on the plaquette chain is performed as explained in Section \ref{sec:LGT}. Since the system is periodic, the subsystem is defined by two boundaries. Their distance defines the subsystem length. Let $N$ denote the number of plaquettes of the periodic chain. Note that due to the boundary links of the subsystem, the subsystem length may take the values $N_A=\{0,...,N+2\}$, where $N_A=1$ and $N_A=N+1$ correspond to only two horizontal links and their complement, respectively. As has been demonstrated analytically and numerically in \cite{Ebner:2024mee}, our computation of entanglement entropy ensures equal results for a subsystem and its complement, i.e., between $N_A$ and $N+2-N_A$.

\begin{figure}[t]
\subfloat[\label{fig:page}]{%
  \includegraphics[width=0.49\linewidth]{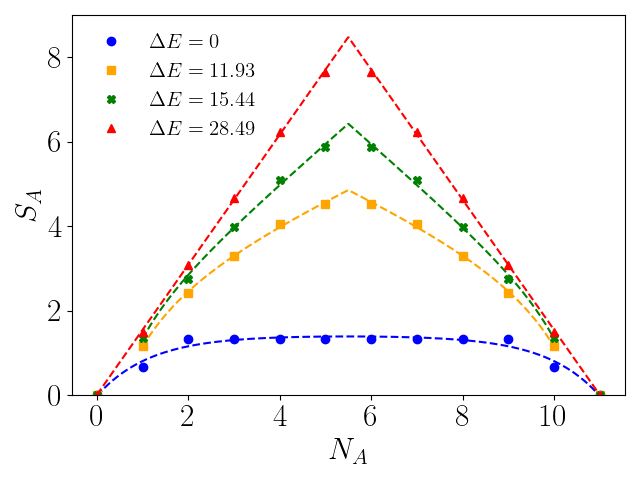}%
}\hfill
\subfloat[\label{fig:all_eigen}]{%
  \includegraphics[width=0.49\linewidth]{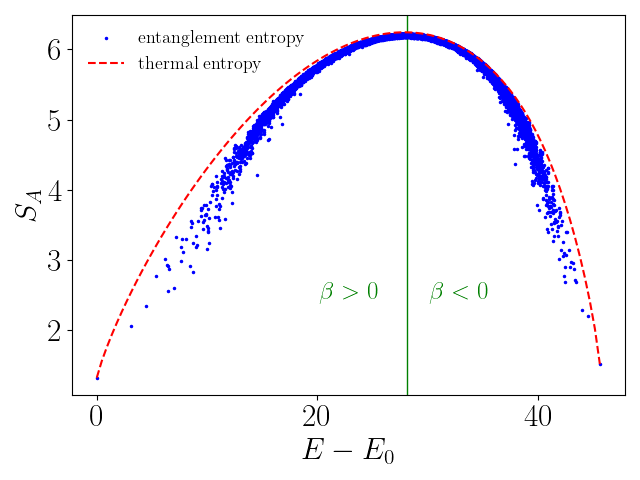}%
}
    \caption{(a) Entanglement entropy as a function of the subsystem size $N_A$ on a periodic nine-plaquette chain in the zero-momentum and reflection symmetric sector with $\jmax=1$. The states are labeled by their energies relative to the ground state energy $\Delta E=E-E_0$ in lattice units. (b) Comparison of entanglement entropy and associated thermal entropy of all eigenstates of the periodic nine-plaquette chain with $\jmax=1$ and $ag^2=1.2$ in a subsystem of length $N_A=4$. The vertical line indicates $\beta=0$ and everything on its left ($\beta>0$) is part of the physical spectrum.}
\label{fig:entanglement}
\end{figure}

Our results for the nine-plaquette chain with $\jmax=1$ and $ag^2=1.2$ are shown in Fig.~\ref{fig:page} for the ground state, two arbitrary excited eigenstates, and the most highly excited eigenstate. The coupling is chosen such that the system exhibits equal partition of electric and magnetic energies, which is a general property of the continuum SU(2) gauge theory in thermal equilibrium.

As expected for a gapped system the ground state satisfies an area law \cite{Hastings:2007iok}. Since the subsystem boundary area is constant, the entanglement entropy quickly saturates to a constant value. Using a three-parameter fit (see Eq.~(10) in \cite{Ebner:2024mee}), we find a ground-state correlation length $\ell_{\rm corr}/a=1.164$. On the other hand, the entanglement entropy for a highly excited state follows a volume law, expressed as a linear dependence on the subsystem length \cite{Ebner:2024mee}. The transition from area to volume law may be expressed by a crossover function that was originally derived for a one-dimensional conformal field theory \cite{Miao:2019xpp}. 
%The use of such an {\it ansatz} for the LGT is motivated by the fact that the continuum SU(2) gauge theory is known to be approximately conformal in the high temperature regime.

This result supports the notion that a pure quantum state cannot be distinguished from a thermal ensemble by observations made on a small subsystem. This is analogous to the resolution of the black hole information ``paradox'' originally proposed by Page \cite{Page:1993wv}: Early in the evaporation process, the Hawking radiation represents a small part of the total system and thus appear thermal until the black hole has lost half its mass. Beyond that point, the entanglement entropy of the subsystem associated with the Hawking radiation steadily decreases, until the full quantum information contained in the black hole becomes visible to an outside observer \cite{Almheiri:2020cfm}. As the results shown in Fig.~\ref{fig:page} are for the nonabelian SU(2) gauge theory, they also support the idea \cite{Ho:2015rga,Muller:2017vnp} that the observed thermal properties of the final state in relativistic heavy ion collisions largely reflect the fact that only a small portion of the final state is experimentally measured.

We further demonstrate this relation by comparing the entanglement entropy of a $N_A=4$ subsystem for all energy eigenstates with their corresponding thermal entropy. Suppose the system is in a pure state $|\psi\rangle$ with energy $E=\langle\psi|H|\psi\rangle$ and define the associated thermal density matrix \cite{Garrison:2015lva}
\begin{linenomath*}
\be
\rho_{A,{\rm th}}(\beta)=\frac{\Tr_{A^c}(e^{-\beta H})}{\Tr(e^{-\beta H})}\label{eq:EE_thermal_RDM}\;,
\ee
\end{linenomath*}
where $\beta$ is chosen such that $E=-(\partial/\partial\beta) \log\,\Tr(e^{-\beta H})$. $\beta=0$ describes the energy for which the density of states has reached its maximum; $\beta=\infty$ corresponds to the ground state. The thermal entropy corresponding to the state $|\psi\rangle$ is then given by the von Neumann entropy of the thermal reduced density matrix shown in Eq.~\eqref{eq:EE_thermal_RDM}.

In Fig.~\ref{fig:all_eigen} we present the entanglement entropy together with the thermal entropy predictions obtained using the thermal reduced density matrix for the $N_A=4$ subsystem of the nine-plaquette chain with $\jmax=1$ in the ergodic coupling regime $ag^2=1.2$. The entanglement entropy of highly excited states in the middle of the spectrum is well described by the thermal prediction, supporting the argument given above.

We draw special attention to the fact that Fig.~\ref{fig:all_eigen} does not show any outliers deviating from the canonical thermal prediction in the middle of the spectrum where the eigenstates form a dense arc. Although such outliers, called quantum many-body scars (QMBS), are found to occur for $\jmax=\frac{1}{2}$, they disappear for higher electric field truncations. Furthermore, at finite lattice size and gauge flux truncation, the existence of QMBS may depend on the coupling $g^2$. Here we choose $ag^2=1.2$ for the nine-plaquette chain with $\jmax=1$ by requiring equipartition of electric and magnetic energies, which is a fundamental property of the continuum SU(2) gauge theory. The plaquette chain, with this coupling and $\jmax=1$, does not exhibit any QMBS, thus supporting the strong ETH for the lattice gauge theory, since for eigenstates up to a given energy, after some critical $j_{\rm max}$, further increasing $j_{\rm max}$ does not change their wavefunctions much~\cite{Turro:2024pxu} and thus the calculated entanglement entropies. For a more detailed discussion, see \cite{Ebner:2024mee}.

\subsection{Time Evolution of Entanglement Entropy}
\label{sec:EE-time}

The main result of the last subsection was the apparent thermal behavior of pure excited states in subsystems. This raises the question how pure quantum states evolve into what appear to be thermal mixtures. In order to answer this question, we next study the real-time evolution of excited pure states of the full system when viewed by observations constrained to a subsystem. The coupling constant is tuned such that the total system is ergodic, i.e., $ag^2=1.2$.

\begin{figure}[t]
\centering
\hfill
\subfloat[Spectrum mean energy state.\label{fig:evo_middle_of_spectrum}]{%
  \includegraphics[width=0.45\linewidth]{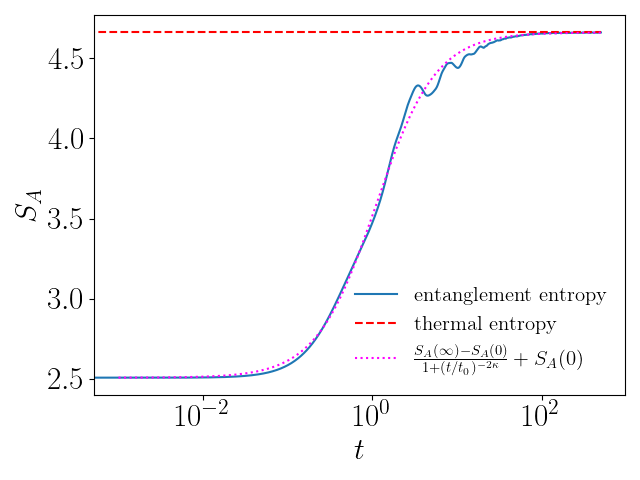}%
}\hfill
\subfloat[Strong coupling ground state.\label{fig:evo_strong_soupling_GS}]{%
  \includegraphics[width=0.45\linewidth]{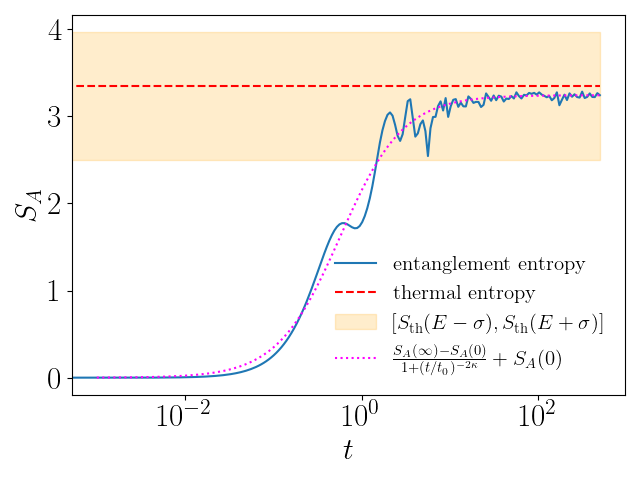}%
}\hfill
\caption{Real-time evolution of entanglement entropy and comparison with the associated thermal entropy for (a) the $k=0$ momentum basis state corresponding to $\beta=0.005$, closest to the spectral mean energy and (b) the strong coupling ground state for a subsystem of length $N_A=3$ on the periodic nine-plaquette chain with $\jmax=1$ and $ag^2=1.2$ in the zero-momentum and reflection symmetric sector. The entanglement entropy growth as a function of time is fitted to a two-parameter function, where we find (a) $\kappa=0.616\pm0.003$, $\log(t_0)=0.097\pm0.004$ and (b) $\kappa=0.608\pm0.014$, $\log(t_0)=-0.555\pm0.021$. The orange region in (b) corresponds to the thermal entanglement entropy of the $1\sigma$-band associated with the energy variance of the strong coupling ground state.}
\label{fig:evo_P9_J1_gsq1.2}
\end{figure}

The first state under investigation is chosen to be a $k=0$ momentum state closest to the mean energy of the spectrum which approximately corresponds to an infinite temperature state ($\beta=0.005\approx 0$). In general, such a state is not an energy eigenstate but a linear combination of all translations and parity transformations of an appropriate electric basis state. Thus, the initial subsystem entanglement entropy of this state is in general non-vanishing. The state evolves under the Hamiltonian of the full system, $|\psi(t)\rangle=e^{-iHt}|\psi(0)\rangle$. The reduced density matrix of the system at time $t$ is given by $\rho_A(t)= {\rm Tr}_{A^c} |\psi(t)\rangle\langle\psi(t)|$, and the entanglement entropy is defined as the von Neumann entropy of $\rho_A(t)$.
Figure~\ref{fig:evo_middle_of_spectrum} shows the time evolution of entanglement entropy of such a momentum eigenstate near the spectral peak in a small subsystem.

For comparison, we study the time evolution of the strong coupling ground state, i.e., the electric basis state whose links are all in the $j=0$ representation of SU(2). At finite coupling, this is not the ground state of the system. Since it is translationally invariant and parity-even, this electric basis state is a $k=0$ momentum eigenstate. Because it is a complete product state, the strong coupling ground state initially has vanishing entanglement entropy on all subsystems. Figure~\ref{fig:evo_strong_soupling_GS} depicts the real-time evolution of the entanglement entropy of this state, starting at $S_A(0)=0$ and reaching a value in the regime of the thermal prediction given by the energy variance $\sigma^2=\langle H^2\rangle-\langle H \rangle^2$ of the state. The corresponding thermal entropy band $[S_{A,\mathrm{th}}(E-\sigma), S_{A,\mathrm{th}}(E+\sigma)]$ for the strong coupling ground state is shown in Fig.~\ref{fig:evo_strong_soupling_GS}.
% For states in the region close to the spectral peak, such as the one in Fig.~\ref{fig:evo_middle_of_spectrum}, this thermal entropy band is narrow.

In order to conduct a more in-depth analysis of the entanglement entropy growth, we fit the time-dependent  entanglement entropy to the function
\begin{linenomath*}
\begin{equation}
S_A(t) = S_A(0)+\frac{S_A(\infty)-S_A(0)}{1+(t/t_0)^{-2\kappa}} \,,
\end{equation}
\end{linenomath*}
containing two parameters $\kappa$ and $t_0$. The fits are shown as dotted pink curves in Fig.~\ref{fig:evo_P9_J1_gsq1.2}. Notably the parameter $\kappa$, which controls the rate of entanglement entropy growth to its asymptotic value, has the same value for both states within the statistical uncertainty. A similar analysis of several other excited states yields a common coupling-dependent value $\kappa = 0.625 \pm 0.020$ suggesting a generic behavior of entanglement entropy growth for excited states.

This two-step entropy evolution has direct implications for the process of entropy production in relativistic heavy-ion collisions, which is thought to be driven by the nonlinear dynamics of SU(3) gauge fields \cite{Fries:2008vp}. First, entanglement entropy is created by the focus on a rapidity window $\Delta y$ that is a small part of the complete collision system which exists in a nearly pure quantum state. The entanglement entropy per unit rapidity $dS_\mathrm{en}/dy$, called decoherence entropy in \cite{Fries:2008vp}, is proportional to the nuclear gluon distribution $xG_A(x)$ at $x=e^{y-y_A}$ where $y_A$ is the beam rapidity \cite{Fries:2008vp,Kharzeev:2017qzs}. (For a critique of this identification, see \cite{Duan:2020jkz}.) This kinematic process of entropy production is followed by the dynamical process of local equilibration or hydrodynamization, which raises the entropy close to its thermal equilibrium value $dS_\mathrm{eq}/dy$. 

The various stages of entropy production and thermalization in relativistic heavy-ion collisions and the role of entanglement in the thermalization process have been studied within a variety of approaches  \cite{Kovchegov:2009du,Muller:2011ra,Huang:2014iwa,Ho:2015rga,Berges:2017hne,Kurkela:2018vqr,Kurkela:2018oqw}.
The overall duration of this process depends on the coupling strength among the gluons in the considered rapidity window. A rough estimate suggests that approximately one quarter of the entropy at the onset of the hydrodynamic stage is due to entanglement between the midrapidity gluons and the rest of the collision system, while half is created during the processes of local equilibration and hydrodynamization. The remainder of the final entropy is generated during the quark-gluon plasma and hadronization stages (see Fig.~2 in \cite{Muller:2011ra}).

\subsection{Bisognano-Wichmann Theorem}
\label{sec:EE-BW}

In previous sections, we showed that the reduced density matrix of a subsystem looks like a thermal state $\rho_A \sim {\rm Tr}_{A^c} (e^{-\beta H})$ at high temperature. In this section, we study the relation between the reduced density matrix and the Hamiltonian for the ground state, which is predicted by the Bisognano-Wichmann (BW) theorem \cite{Bisognano:1975ih,Bisognano:1976za,Headrick:2019eth}. We define the modular Hamiltonian $\mathcal{H}_A$, also called the entanglement Hamiltonian, as $\rho_A \equiv e^{-\mathcal{H}_A}/\mathrm{Tr}(e^{-\mathcal{H}_A})$. The original BW theorem states that the modular Hamiltonian associated with the half space (say, $x_1>0$) for a relativistic quantum field theory is given by
\begin{linenomath*}
\begin{align}
\label{eqn:HA}
\mathcal{H}_A = \int_0^{\infty} d x_1 \beta(x_1) H(x_1)\,,
\end{align}
\end{linenomath*}
where $\beta(x_1)=\beta x_1$ for some constant $\beta$ and $H(x_1)$ denotes the Hamiltonian density. The expression~\eqref{eqn:HA} has been generalized for a finite interval in ($1+1$)-dimensional conformal field theories \cite{Casini:2011kv,Cardy:2016fqc} and for lattice theories \cite{Giudici:2018izb,Mueller:2021gxd} with open and periodic boundary conditions. The reduced density matrix of the ground state can be interpreted as a thermal density matrix with locally varying temperature. For periodic boundary conditions $\beta(x)$ shrinks when the boundary of the subsystem is approached. 

\begin{figure}[t]
\centering
\subfloat[\label{fig:BW_spin}]{%
  \includegraphics[width=0.45\linewidth]{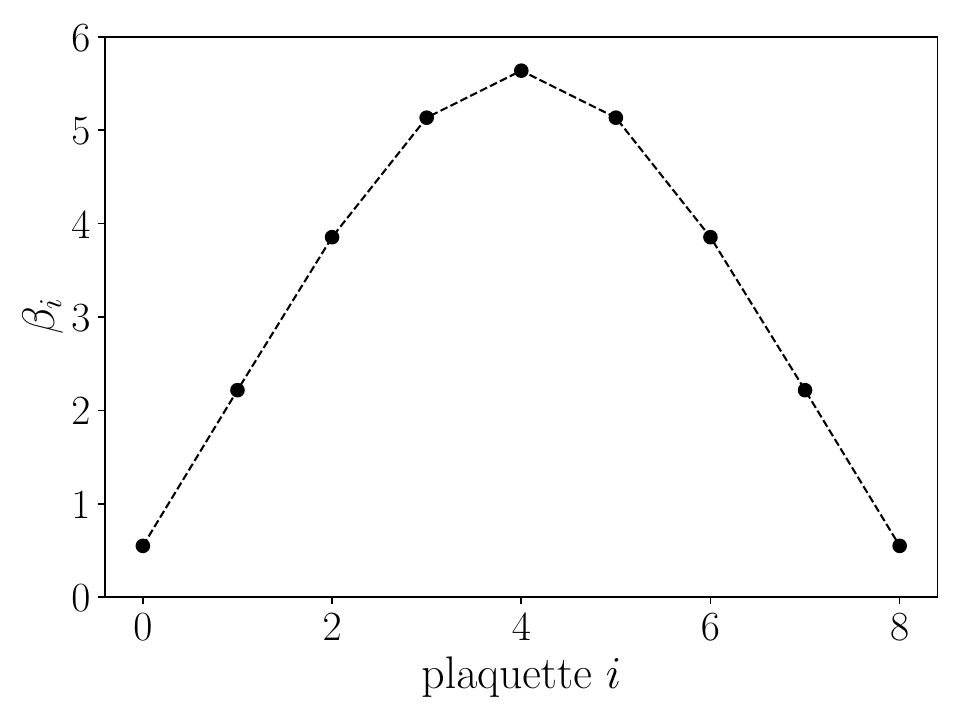}%
}\hfill
\subfloat[\label{fig:beta_profile_fit_all_PB_N9_0_5_g2_1_jmax1}]{%
  \includegraphics[width=0.45\linewidth]{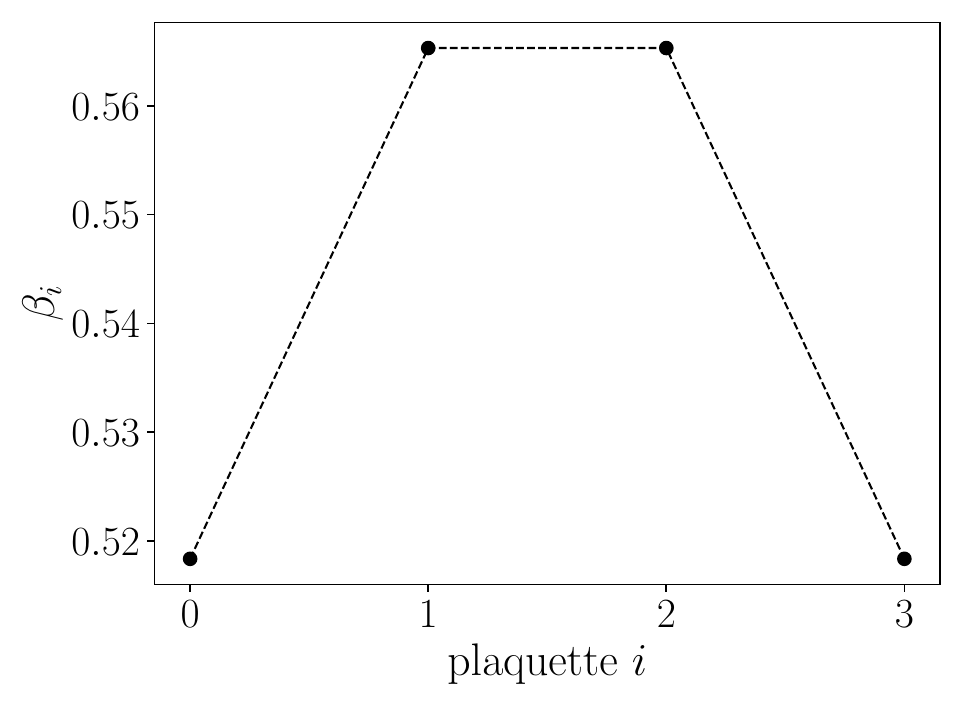}%
}\hfill
\caption{Inverse temperature profile $\beta_i$ for the optimized Bisognano-Wichmann approximation $\mathcal{H}_A^{\rm BW}$ to the lattice modular Hamiltonian for plaquettes with (a) $N=19$, $\jmax=\frac{1}{2}$ and periodic boundary conditions and (b) $N=9$, $\jmax=1$ and periodic boundary conditions in the zero-momentum and reflection symmetric sector. 
Note that $\beta_i$ approaches zero at the boundaries of the interval for case (a), but not for (b).}
\label{fig:BW-temp}
\end{figure}
Here we study the modular Hamiltonian associated with a half chain ($N_A=9$) on a periodic plaquette chain with $N=19$, $\jmax=\frac{1}{2}$ and $ag^2=1.2$ in the Ising representation of the gauge theory. We write the lattice BW Hamiltonian as
$\mathcal{H}_A^{\rm BW} = \sum_{i\in A} \beta_i H_i$, where the single-site Hamiltonian is defined in Eq.~\eqref{eq:H_Ising}. Since the SU(2) gauge theory is not conformal, we treat $\beta_i$ as parameters, which are chosen to minimize the Frobenius norm of $\rho_A - \rho_A^{\rm BW}$, where
$\rho_A^{\rm BW} = e^{-\mathcal{H}_A^{\rm BW}}\!/{\rm Tr}(e^{-\mathcal{H}_A^{\rm BW}})$. We note that the modular Hamiltonian is independent of the boundary condition of the subsystem in the known examples that have been analyzed in the continuum limit, where $\beta(x) \to 0$ on the boundary. However, a site on a lattice cannot be arbitrarily close to the boundary due to the lattice spacing. Thus we expect the lattice modular Hamiltonian to show some dependence on the boundary condition. Here we choose the averaged boundary condition for the spin chain, i.e., $\sigma_i^z\to0$ if $i$ is just outside the boundary. The $\beta_i$ profile optimized for the half chain $N_A=9$ is shown in Fig.~\ref{fig:BW_spin}. The optimized Frobenius norm is $\|\Delta\rho_A\|_F \equiv \|\rho_A - \rho_A^{\rm BW}\|_F \approx 0.0215$. This is much smaller than the Frobenius norm of the reduced density matrix itself, $\|\rho_A \|_F \approx 0.8836$ and indicates that the BW form of the entanglement Hamiltonian with optimized parameters $\beta_i$ closely approximates the true entanglement Hamiltonian. 

Proceeding from the crude approximation of the LGT formulated as spin chain to the case $\jmax=1$, the reduced density matrix and thus also the modular Hamiltonian breaks up into different blocks corresponding to the different dangling links at the cut boundary. In the following we consider the half partition ($N_A=\lfloor \frac{N}{2}\rfloor$) of a nine-plaquette chain with periodic boundary conditions. $\rho_A$ breaks up into 25 blocks. As before, we will not take into account that the $\beta_i$ profile can depend on the cut boundary condition, but minimize the Frobenius norm of the difference between the reduced density matrix and $\rho_A^{\rm BW}$ with the same fit parameters $\beta_i$ for all cut boundary conditions. We are able to minimize the Frobenius norm to $\|\Delta\rho_A\|_F \approx 0.07663$ ($||\rho_A||_F=0.5204$) with the $\beta_i$ profile shown in Fig. \ref{fig:beta_profile_fit_all_PB_N9_0_5_g2_1_jmax1}.
The von Neumann entropies of the ground state calculated with $\rho_A^{\rm BW}$ and $\rho_A$ agree within 4\% and in both cases the temperature $T_i=1/\beta_i$ is highest at the boundary.

\begin{figure}[t]
\centering
\subfloat[\label{fig:BW_eigenvalue0.5}]{%
  \includegraphics[width=0.45\linewidth]{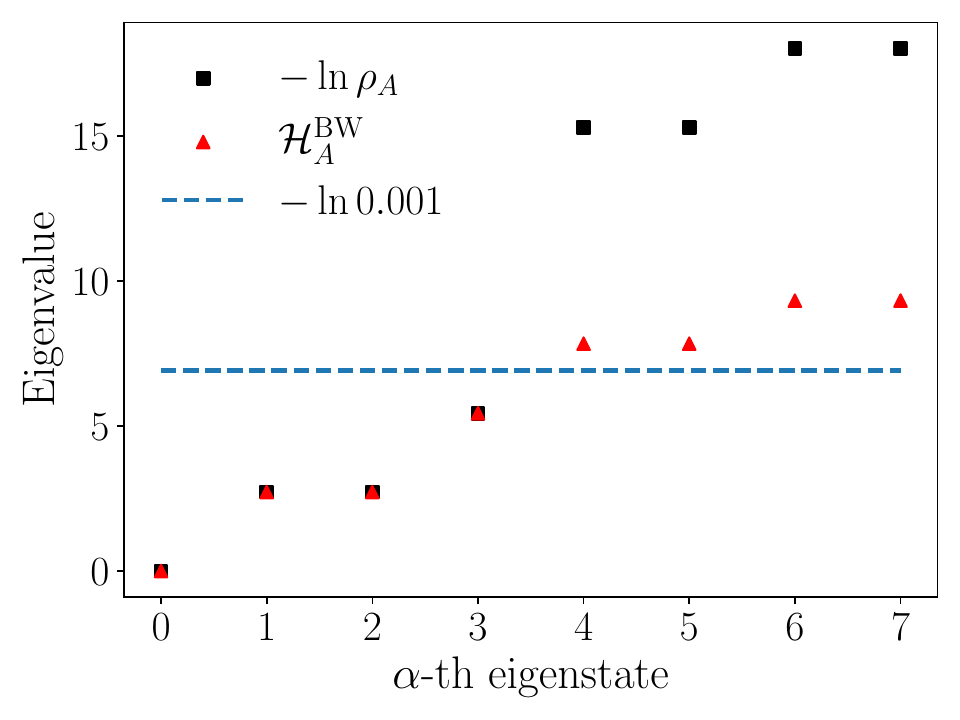}%
}\hfill
\subfloat[\label{fig:BW_eigenstate0.5}]{%
  \includegraphics[width=0.46\linewidth]{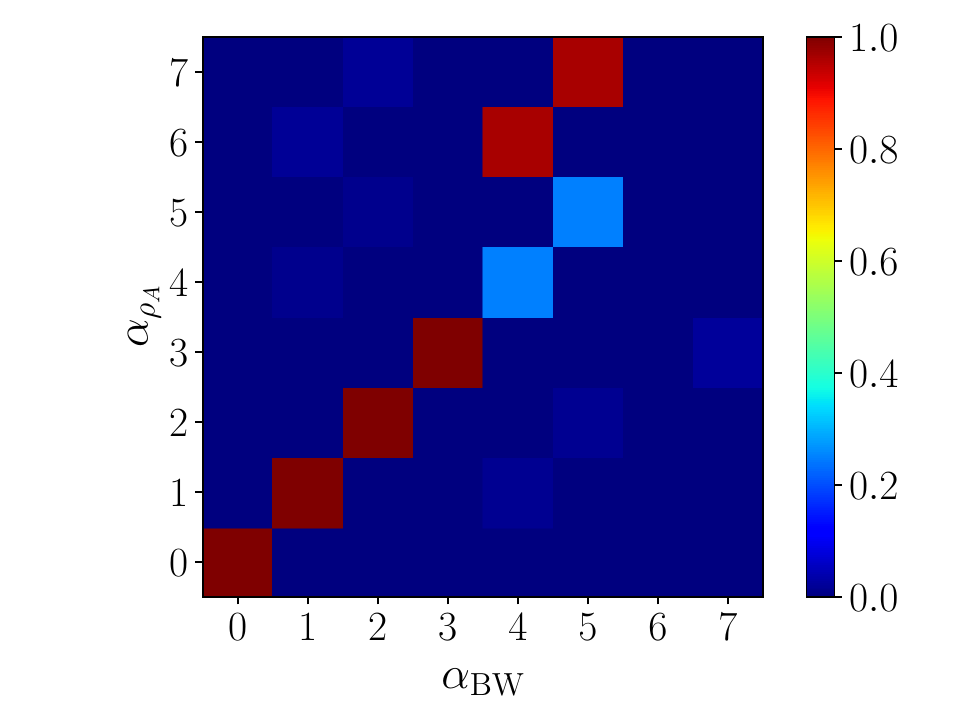}%
}

\subfloat[\label{fig:BW_eigenvalue1}]{%
  \includegraphics[width=0.45\linewidth]{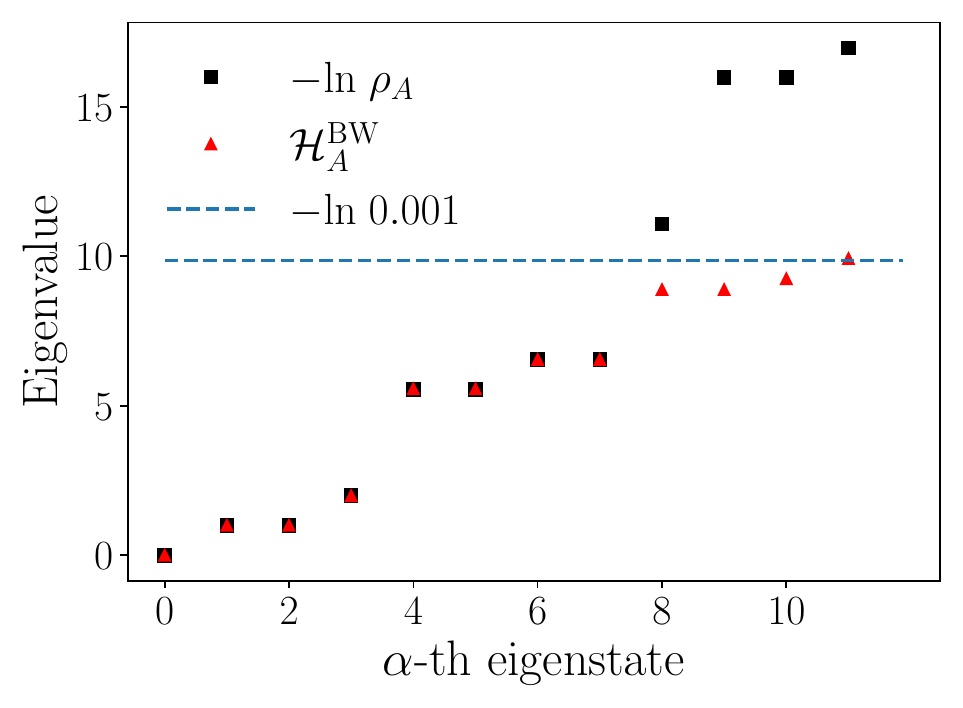}%
}\hfill
\subfloat[\label{fig:BW_eigenstate1}]{%
  \includegraphics[width=0.46\linewidth]{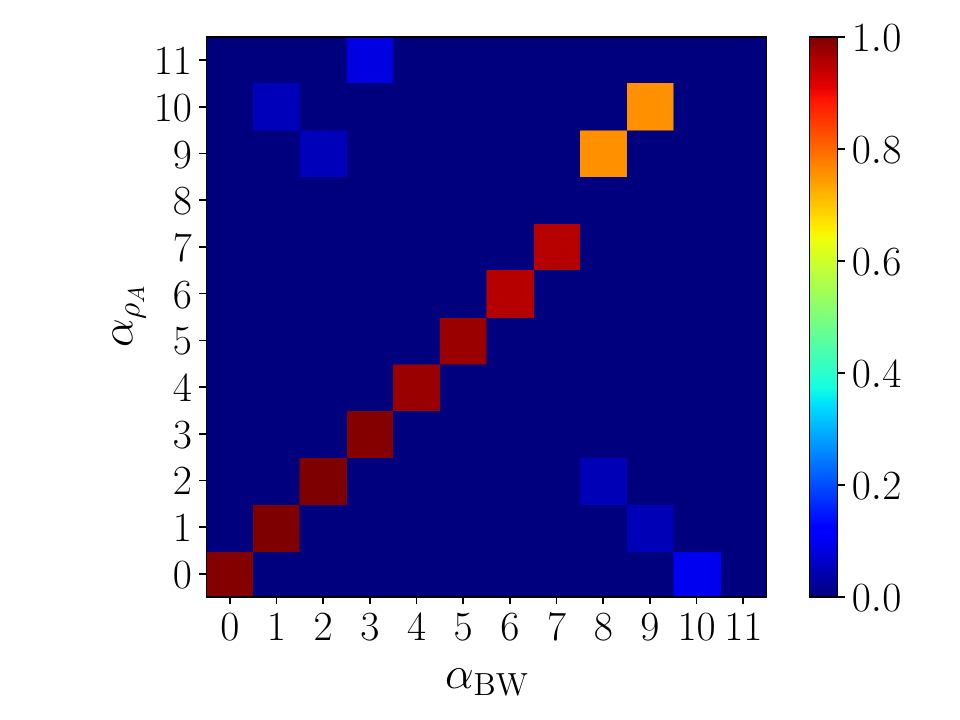}%
}
\caption{Comparison of the entanglement spectra and the spectra of the optimized modular Hamiltonian (left) and the eigenstate overlaps $|\langle \psi_{\alpha_{\rho_A}} | \psi_{\alpha_{\rm BW}}\rangle|$ (right) for the $N=19$, $j_{\rm max}=\frac{1}{2}$ chain (top) and the nine-plaquette chain with $j_{\rm max}=1$ (bottom).}
\label{fig:entanglement_spectra}
\end{figure}

To demonstrate the quality of the optimization visually, we compare the entanglement spectra (eigenvalues of $-\ln\rho_A$) and the spectra (eigenvalues) of the optimized modular Hamiltonian $\mathcal{H}_A^{\rm BW}$, as well as the corresponding eigenstates. The results are shown in Fig.~\ref{fig:entanglement_spectra}. For the comparison of the spectra, we shift them such that the lowest eigenvalue is zero for both cases. The two spectra agree well for the low-lying states and large deviations are only seen for eigenvalues above $-\ln0.001\approx 6.91$, whose contribution to the Frobenius norm is less than $0.1\%$. Optimization based on the Frobenius norm is not sensitive to these high-lying states, and thus the eigenvalues of $-\ln\rho_A$ and the optimized $\mathcal{H}_A^{\rm BW}$ can be very different in this range. The low-lying eigenstates of $-\ln\rho_A$ have a one-to-one correspondence with those of $\mathcal{H}_A^{\rm BW}$, which can be seen from the eigenstate overlaps, i.e., $|\langle \psi_{\alpha_{\rho_A}} | \psi_{\alpha_{\rm BW}}\rangle|$, as shown in the right column of Fig.~\ref{fig:entanglement_spectra}. For high-lying eigenstates with spectra above $6.91$, the correspondence is broken, which has almost no effect on the Frobenius norm.

\subsection{Anti-Flatness of the Entanglement Spectrum}
\label{sec:magic}

The structure of the entanglement spectrum, which is obtained by diagonalizing the reduced density matrix, contains important information about the quantum complexity of the state. 
An easily calculated measure of this ``quantumness'' or ``magic'' is the rescaled lack of flatness (anti-flatness) of the entanglement spectrum (see e.g.\ Eq.(4) in \cite{Tirrito:2023fnw}):
\begin{linenomath*}
\begin{align}\label{eq:12}
M_A := \frac{{\cal F}_A(\rho_A)}{[{\rm Tr}(\rho_A^2)]^2}= \frac{{\rm Tr}(\rho_A^3) - [{\rm Tr}(\rho_A^2)]^2}{[{\rm Tr}(\rho_A^2)]^2} = \frac{{\rm Tr}(\rho_A^3)}{[{\rm Tr}(\rho_A^2)]^2} - 1\,.
\end{align}
\end{linenomath*}
Many definitions for magic have been proposed in the literature. Here, we rely on \cite{Cao:2024nrx} where it was shown that anti-flatness provides a lower bound for ``non-local magic" which, in turn, bounds the hardness  of classical simulations from below. The precise relationship between different definitions of magic and the impact on computational hardness is subtle. We refer to \cite{Cao:2024nrx} for details and references. 
It has been shown that for highly entangled states one only needs to average anti-flatness over a few Clifford orbits to obtain an estimate of magic~\cite{Tirrito:2023fnw}. This is a relevant result, as our main interest here are highly excited states that are also highly entangled, with the goal to better understand the quantum complexity of, e.g., real-time simulations of thermalization in SU($N$) gauge theories.
As a first step, we focus here on the normalized anti-flatness $M_A$ in the entanglement spectrum for SU(2) gauge theory. 

\begin{figure}[t]
\centering
\hfill
\subfloat[\label{fig:tr_rho2}]{%
  \includegraphics[width=0.33\linewidth]{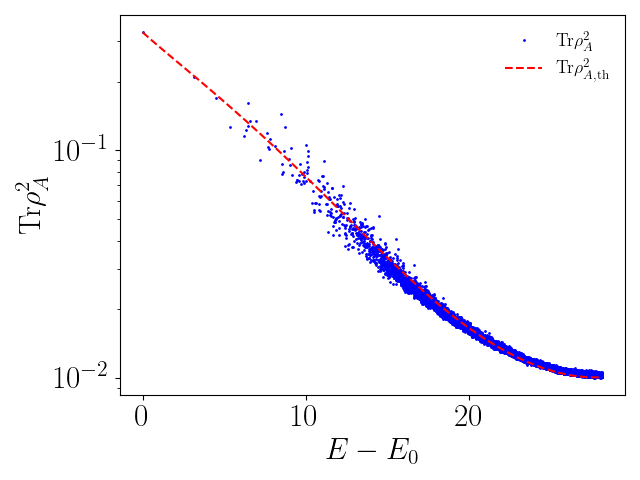}%
}\hfill
\subfloat[\label{fig:tr_rho3}]{%
  \includegraphics[width=0.33\linewidth]{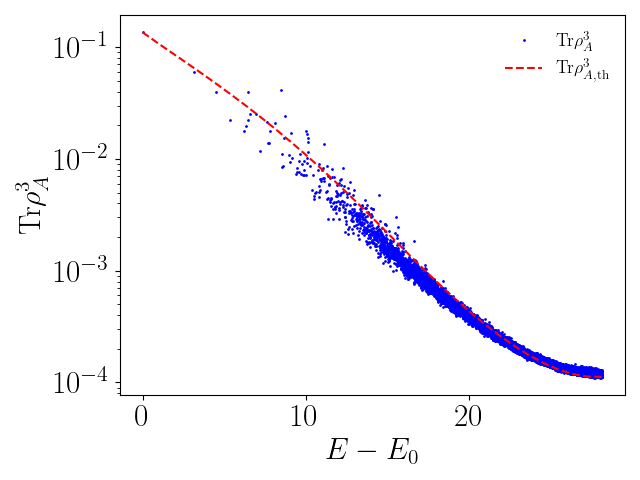}%
}\hfill
\subfloat[\label{fig:magic_spec_NA3}]{%
  \includegraphics[width=0.33\linewidth]{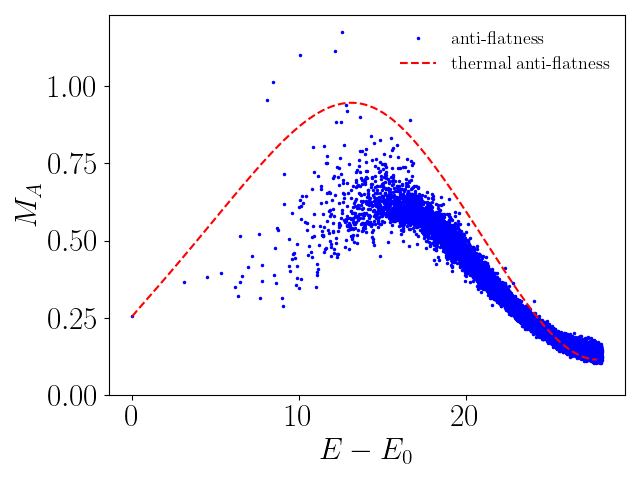}%
}\hfill
\caption{The blue dots show
(a) ${\rm Tr}\rho_A^2$,
(b) ${\rm Tr}\rho_A^3$,
and (c) anti-flatness of the entanglement spectrum $M_A$ of all energy eigenstates below the spectral peak (positive  temperature states) in the zero-momentum and reflection symmetric sector on a periodic nine-plaquette chain with $j_{\rm max}=1$, $ag^2=1.2$ for subsystem length $N_A=3$. For comparison, the dashed red line shows the respective prediction for the reduced thermal density matrix $\rho_{A,\rm th}$ on the same chain.}
\label{fig:thermal_magic}
\end{figure}

Results for the physical part of the spectrum (the region of increasing spectral density $d\rho(E)/dE >0$ corresponding to $\beta>0$ that contains approximately 25,000 states) of the periodic nine-plaquette chain with $j_{\rm max}=1$ in the ergodic coupling regime ($ag^2=1.2$) are shown in Fig.~\ref{fig:thermal_magic}.
The subsystem length is $N_A=3$ and we analyze the zero-momentum and reflection symmetric sector.

In Fig.~\ref{fig:tr_rho2} we show ${\rm Tr}\rho_A^2$ which agrees perfectly with the thermal prediction based on the reduced thermal density matrix \eqref{eq:EE_thermal_RDM}.
Figure~\ref{fig:tr_rho3} is the corresponding plot for ${\rm Tr}\rho_A^3$ which shows a slight systematic deviation between the exact and thermal values. This indicates that the true third-order R\'enyi entropy of the subsystem reveals quantum correlations that are not present in the canonical ensemble. 
Figure~\ref{fig:magic_spec_NA3} shows the anti-flatness measure $M_A$ defined in Eq.(\ref{eq:12}). Remarkably, while ${\rm Tr}(\rho_A^3)$ and ${\rm Tr}(\rho_A^2)$ are close to their thermal predictions, $M_A$, which is given by the ratio in Eq.(\ref{eq:12}), is not. This suggests strong correlations that are different from Gaussian and thermal distributions.
This finding conforms to the general argument that magic and other observables based on higher R\'enyi entropies provide more detailed information about the entanglement structure of a quantum system than are revealed by the entanglement entropy
\cite{Dong:2016fnf}.

\begin{figure}[t]
\centering
\includegraphics[width=0.67\linewidth]{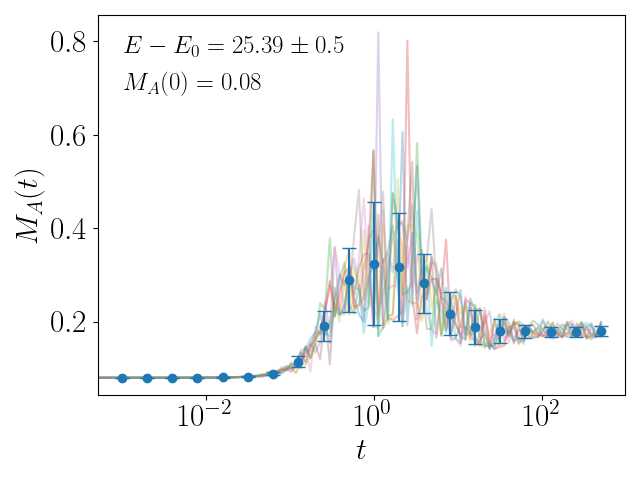}
\caption{
Real-time evolution of the anti-flatness for an ensemble of 200 excited states in the energy window $E-E_0=25.39\pm0.5$ on the periodic nine-plaquette chain with $j_{\rm max}=1$, $ag^2=1.2$ for a subsystem of length $N_A=3$. The states are chosen such that at $t=0$ they all have $M_A(0)=0.08$.
The blue dots and error bars represent the average and standard deviation of the whole ensemble. The colored lines depict the time evolution for ten arbitrarily chosen states.
}
\label{fig:magic}
\end{figure}

As a next step, we investigate the time evolution of anti-flatness during thermalization. The energy window for our study is chosen near the upper end of the physical range where the eigenstate anti-flatness approximately matches the thermal anti-flatness. The initial states are chosen as coherent superpositions of eigenstates within an energy window $\Delta E = 1$.
The real-time evolution of the normalized anti-flatness of these excited states is depicted in Fig.~\ref{fig:magic}. The figure shows that the time window $t/a \sim 1$, which is characterized by a rapid rise in the entanglement entropy (see Fig.~\ref{fig:evo_P9_J1_gsq1.2}), is associated with high anti-flatness that decreases again as the subsystem density matrix approaches equilibrium.
As is also evident, the time evolution of this ``magic barrier'' (rather than the more exact anti-flatness barrier) varies widely and erratically for different initial states, resulting in a rather large variance around the ensemble average. In order to circumvent this erraticity, we quantify the magic barrier of each individual state by its peak anti-flatness divided by the initial anti-flatness, ${\max}_t\left\{M_A(t)/M_A(0)\right\}$. For the ensemble of the 200 states used in Fig.~\ref{fig:magic}, we obtain a value of $7.34\pm1.50$, giving an almost $5\sigma$ signal.
Our results suggest that simulations of equilibration dynamics demand full quantum computing, in contrast to thermal equilibrium properties that can be studied efficiently using imaginary time techniques. 
Clearly, the phenomenology of magic for non-stationary states deserves further study. Hopefully, this will lead to a better understanding of features in the thermalization process that cannot be appropriately captured by kinetic theory.

\subsection{Toward Quantum Computation}
\label{sec:QC}

Quantum computers implement the unitary evolution of a quantum state by means of the successive application of quantum gates onto the initial state. By directly simulating the unitary time evolution for a specific state they avoid the need to find the eigenstates of the Hamiltonian in a chosen basis.
A drawback of this approach is that it is impractical to demonstrate the validity of the ETH for a quantum system. One therefore has to consider other methods to study thermalization. 

One general method is to start with a well-chosen initial state, simulate its time evolution and observe the approach of the expectation value of an operator of interest to its thermal value
\begin{linenomath*}
\begin{equation}
\mathcal{A}_\Psi(t) = \langle\Psi(t) |\mathcal{A}| \Psi(t)\rangle \underset{t\to\infty}{\longrightarrow} \mathcal{A}_{\rm th}\,.
\end{equation}
\end{linenomath*}
The existence of an intermediate hydrodynamic regime governing the dynamics immediately preceding full thermalization will imply that most operators will exhibit similar behavior before reaching their thermal values. For local operators $\mathcal{A}(x)$, this behavior should be expressed in a common space-time variation of their expectation values $\mathcal{A}_{\rm th}(x,t)$ representing local, but not global, thermal equilibrium. It is believed that this property is encoded in the plateau structure of the spectral function $f_\mathcal{A}(E,\omega)$ in the fundamental ETH relation (\ref{eq:ETH}) for small $\omega$. In the absence of conservation laws other than that of energy and momentum, generic operators are expected to exhibit the same diffusive plateau.

From the quantum algorithm perspective, this study can be performed by starting with an easy-to-prepare initial state such as a product state, evolving the state in time by Trotterizing the Hamiltonian evolution, and finally measuring the state in a basis where the operator $\mathcal{A}$ is diagonal. With Trotterization and local Hamiltonians as in gauge theories, it is expected that quantum simulation of the Hamiltonian time evolution only requires resources that are polynomial in accuracy and lattice size, though it is not clear how to optimally map local gauge degrees of freedom onto qubits in general. One can also use entanglement entropy as an observable to study thermalization as demonstrated in Section~\ref{sec:EE-time}. For a small subsystem, its reduced density matrix can be reconstructed by quantum state tomography, and then the entanglement entropy can be calculated. For larger subsystems, one can use the randomized measurement toolbox~\cite{Elben:2022jvo}, which has been used for entanglement entropy calculations. A recent experimental study has shown the area and volume laws in a simple Hamiltonian system that maps naturally onto the quantum device (analog simulation)~\cite{Andersen:2024aob}. For an initial state that mimics heavy-ion collisions, it is expected that confinement phenomena (hadronization) become important during the hydrodynamic stage before the full thermalization. It is thus interesting to explore how and when this effect manifests itself in the real-time evolution.

Another way of using quantum computing to study thermalization is to calculate real-time correlation functions of local operators at thermal equilibrium, such as the retarded or Wightman correlation function of the stress-energy tensor. From the perspective of linear response theory, this way is equivalent to the first method discussed above. Hydrodynamic behavior toward global equilibrium can be studied at late time~\cite{Arnold:1997gh}. Transport coefficients can also be extracted with the help of the Kubo formula~\cite{Turro:2024pxu}. From the quantum algorithm perspective, this calculation starts with preparing a thermal state, applies a perturbation given by the local operator, then evolves in time and finally measures the local operator. The thermal state preparation is a well known and much discussed problem, see, e.g.,~\cite{Temme:2009wa,Chen:2023cuc} for recent references. We can use the quantum imaginary time propagation method~\cite{Turro:2023nuk}, also employed in~\cite{Turro:2024pxu} which becomes costly at low temperature and large volume. Since we are mainly interested in thermalization at high temperature, e.g., in the deconfined phase of QCD matter, we expect this to be a less limiting factor. Nevertheless, another thermal state preparation method is based on the Lindblad equation, whose evolution time has been shown to scale only polynomially with the system size in the case of the Schwinger model~\cite{Lee:2023urk}. One can further explore whether different local operators have the same thermalization time scale in this way.

\section{Summary and Discussion}
\label{sec:SUM}
 
Motivated by the parallel progress of theoretical and experimental investigations of increasing sophistication and complexity that all confirm the validity of the Standard Model, the focus has shifted to those aspects of gauge theories that cannot be understood perturbatively. One such broad category includes the dynamics of thermalization studied experimentally in high-energy heavy-ion collisions. Rigorous answers to all open questions related to thermalization require fundamental theoretical progress on many fronts including, e.g., quantum computing, holographic techniques, and high-performance digital computing for tensor networks and real-time lattice gauge theory. We are presently witnessing just the beginning of these developments. 

In this review we reported a few recent results for one subfield, namely digital Hamiltonian calculations for the simplest fundamental case, pure SU(2) gauge theory, on small lattices. There already exists a large body of results, primarily from the study of simple numerical models that provide guidance for future investigations. As these are far too many to give a fair appraisal of all of them, we limit ourselves to just citing a few very recent ones which we found especially stimulating for our own work. These also provide many references to earlier relevant publications. \cite{Kavaki:2024ijd,Babusci:2024lgx,Kadam:2024zkj,Grabowska:2024emw,Ale:2024uxf,Zhang:2024fgv,Farrell:2024mgu,Kavaki:2025hcu,Patil:2025ump,Yang:2025edn}. Their results concern, e.g., the validity of the ETH, which has been systematically investigated and elementary aspects of the entanglement properties of gauge theories. A one-sentence summary is that, within numerical accuracy and accessible lattice size, pure SU(2) gauge theory fulfills all expectations for an information scrambling and rapidly thermalizing quantum system investigated so far, but that there is much more work to be done.

The ETH posits that the matrix elements of generic local operators between energy eigenstates have the particular form stated in Eq.~(\ref{eq:ETH}). The ETH asserts certain random matrix properties, which have been checked and confirmed by exact diagonalization of the Hamiltonian in Hilbert spaces of dimension up to $O(10^6)$. The numerical simulations reviewed in Section~\ref{sec:ETH} not only confirm the GOE properties posited by ETH, but also permit to determine the time scales on which they are established. When extended in the future to larger lattices by more powerful numerical methods, these time scales will provide phenomenologically valuable results for the short-lived QCD fireballs created in high-energy heavy-ion collisions. 

In Section~\ref{sec:SFF} we presented, for the first time, evidence that on long time scales the spectral form factor for the SU(2) LGT exhibits the slope-ramp-plateau structure expected for a chaotic quantum system. The transition between the slope and ramp regions allows us to extract the diffusion constant for the studied system, another phenomenologically relevant quantity. In Section~\ref{sec:EE-Page} we presented the Page curve for the entanglement entropy of a subsystem, which agrees perfectly with analytic predictions, including the transition from area to volume scaling for increasing energy of the eigenstate. In contrast to studies of other many-body quantum systems we find no indication for quantum many-body scars for larger values of $j_{\rm max}$ and plaquette number $N$, a finding that is relevant for the continuum limit. In Section~\ref{sec:EE-time} we analyse the time evolution of Entanglement Entropy. 
Furthermore, we presented an analysis of the Bisognano-Wichmann theorem (Section~\ref{sec:EE-BW}) for the entanglement Hamiltonian of the LGT and determined the optimal temperature profiles. Finally, we calculated the anti-flatness of the entanglement spectrum, which provides a lower bound for magic (Section~\ref{sec:magic}).
The real-time evolution of non-eigenstates reveals a ``magic barrier'' within equilibration dynamics, underscoring the necessity of full quantum computation for efficiently simulating thermalization in nonabelian gauge theory. 

Altogether, the material presented here constitutes reassuring progress toward the goal of establishing entanglement and thermalization properties of the nonabelian gauge field described in the beginning. We need to keep in mind, however, that these results were obtained for only the $(2+1)$-dimensional pure SU(2) gauge theory on rather small lattices. The extension to $(3+1)$ dimensions, other gauge groups, and larger lattices are obvious near future goals, which will likely require quantum computing. Fortunately, quantum algorithms for gauge theories have seen significant progress in recent years and it is hoped that quantum computing will become a standard tool in the intermediate-term future. These and other numerical techniques have already begun to become genuine discovery tools. In contrast, the inclusion of fermions is probably rather a mid-future goal as it requires us to first understand complications like the potential appearance of quantum many-body scars~\cite{Halimeh:2022rwu,Desaules:2023yhw,Calajo:2024bvs}.

\section{Data availability statement}
\label{sec:Data}

The data used in this work can be reproduced using a custom code (see the Code availability statement).

\section{Code availability statement}
\label{sec:Code}
We will make our code available to everybody interested, both the version which was used for this paper and future improved versions. To this end we will create a persistent repository once this manuscript got accepted for publication and before it gets published. We are also happy to discuss specific aspects of our work in greater detail. If you are interested, please contact andreas.schaefer@physik.uni-r.de.

\bibliography{main}% common bib file
%% if required, the content of .bbl file can be included here once bbl is generated
%%\input NAME.bbl

\vskip 1 cm 
{\bf Acknowledgments:}

The authors gratefully acknowledge the scientific support and HPC resources provided by the Erlangen National High Performance Computing Center (NHR@FAU) of the Friedrich-Alexander-Universit\"at Erlangen-N\"urnberg (FAU) under the NHR project b172da-2. NHR funding is provided by federal and Bavarian state authorities. NHR@FAU hardware is partially funded by the German Research Foundation (DFG 440719683). BM acknowledges support from the U.S. Department of Energy, Office of Science, Nuclear Physics (awards DE-FG02-05ER41367). XY is supported by the U.S. Department of Energy, Office of Science, Office of Nuclear Physics, InQubator for Quantum Simulation (IQuS) (https://iqus.uw.edu) under Award Number DOE (NP) Award DE-SC0020970 via the program on Quantum Horizons: QIS Research and Innovation for Nuclear Science.

\vskip 1 cm
{\bf Author contributions:}

Lukas Ebner, Clemens Seidl, and Xiaojun Yao were in charge of code development and run submission at NHR@FAU for subsections \ref{sec:ETH}, \ref{sec:EE-Page}, \ref{sec:EE-time}, \ref{sec:EE-BW}, and \ref{sec:magic}, often splitting the task for plaquette chains and honeycomb lattices between them. Leonhard Schmotzer was in charge of code development and run submission for subsections \ref{sec:SFF} and \ref{sec:EE-BW}. Berndt M\"uller, Andreas Sch\"afer and Xiaojun Yao provided subproject ideas and some needed technical know-how. Also, they jointly supervised the work. All authors contributed to the analyses and discussion of results for all subsections.

\vskip 1 cm
{\bf Competing interests:}

All authors declare no competing interests

\end{document}